\documentclass[11pt]{article}

\usepackage{amsfonts, amssymb}
\usepackage{amsmath}
\usepackage{verbatim}
\usepackage{indentfirst}
\usepackage{color}

\begin{document}

\date{}
\title{{\large \bf The Classification of  $n$-Lie
Algebras
\thanks{ \small {Project partially supported by NSF(10871192) of
China, NSF(A2010000194) of Hebei Province, China.  Email address:
bairp1@yahoo.com.cn; songguojie821209@yahoo.com.cn;
yzz@maths.uq.edu.au }}}}
\author{\small Rui-pu  Bai$^{1}$  ~Guo-jie Song$^1$ ~Yao-zhong
Zhang$^2$
\\ \small 1. College of Mathematics and Computer,
\\\small Key Lab. in Machine Learning and Computational
Intelligence,
\\\small
Hebei University, Baoding (071002), China
\\\small 2. School of Mathematics and Physics, The University of Queensland,
\\\small Brisbane, QLD 4072, Australia}

\maketitle

\noindent{ \bf Abstract:} ~This paper proves the isomorphic
criterion theorem for $(n+2)$-dimensional $n$-Lie algebras, and
gives a complete classification of $(n+1)$-dimensional $n$-Lie
algebras and $(n+2)$-dimensional $n$-Lie algebras over an
algebraically closed field of characteristic zero.


 \vspace{2mm} \noindent{\bf 2000 MR subject
classification}: ~17B05 ~17D99

\vspace{2mm}\noindent{ \bf 1. Introduction}

\vspace{2mm} In 1985, Filippov [3] introduced the concept of $n$-Lie
algebras and classified the $(n+1)$-dimensional $n$-Lie algebras
over an algebraically closed field of characteristic zero. The
structure of $n$-Lie algebras is very different from that of Lie
algebras due to the $n$-ary multilinear operations involved. The
$n=3$ case, i.e. 3-ary multilinear operation, first appeared in
Nambu's work [1] in the description of simultaneous classical
dynamics of three particles. In that work, Nambu extended the
Poisson bracket and arrived at the generalized Hamiltonian equation
involving a 3-ary multilinear bracket
 $\{ \, ,  \,  , \}$.
Takhtajan [2] investigated the geometrical and algebraic aspects of
the generalized Nambu mechanics, and established the connection
between the Nambu mechanics and Filippov's theory of $n$-Lie
algebras [3].

The development of $n$-Lie algebras has opened a new chapter in the
study of Lie theory, attracting much attention in different research
areas due to their close connections with dynamics, geometries as
well as string and membrane theories. For example, Bagger and
Lambert [4] proposed a field theory model for multiple M2-branes
based on the metric $n$-Lie algebras, and the authors in [5] found
new $3$-Lie algebras and their applications in membranes. More
applications of the $n$-Lie algebras can be found in  [6, 7, 8, 9,
10, 11, 12, 13].

It is known that up to isomorphisms there is a unique finite
dimensional simple $n$-Lie algebra for $n>2$ over an algebraically
closed field of characteristic zero [14], which is the
$(n+1)$-dimensional $n$-Lie algebra.
So far, the only known infinite dimensional simple $n$-Lie algebras
over fields of characteristic $p\geq 0$ are Jacobian algebras and
their quotient algebras [15, 16]. The first author of the current
paper and her collaborators [17] showed that there exist only
$[\frac{n}{2}]+1$ classes of $(n+1)$-dimensional simple $n$-Lie
algebras over a complete field of characteristic $2$. They also
showed that there are no simple $(n+2)$-dimensional $n$-Lie
algebras.

In [22], $6$-dimensional $4$-Lie algebras were classified and some
basic properties of  $(n+2)$-dimensional $n$-Lie algebras were
studied. The purpose of this paper is to classify the
$(n+2)$-dimensional $n$-Lie algebras over an algebraically closed
field of characteristic zero. Our results are expected to be useful
in various applications.

The organization for the rest of this paper is as follows. Section
$2$ introduces some basic notions. Section $3$ is devoted to the
properties and classification of the $(n+2)$-dimensional $n$-Lie
algebras.

\vspace{2mm}\noindent{ \bf 2. Fundamental notions}

An $n$-Lie algebra is a vector space $A$ over a field $F$
($char(F)\neq 2$) equipped
 with an $n$-multilinear operation $[x_1, \cdots, x_n]$ satisfying
$$ [x_1, \cdots, x_n]=sgn(\sigma)[x_{\sigma (1)}, \cdots, x_{\sigma(n)}], \eqno(2.1) $$  and
 $$
  [[x_1, \cdots, x_n], y_2, \cdots, y_n]=\sum_{i=1}^n[x_1, \cdots, [ x_i, y_2, \cdots, y_n], \cdots, x_n] \eqno(2.2)
$$
for any $x_1, \cdots, x_n, y_2, \cdots, y_n\in A$ and any
permutation $\sigma\in S_n.$  Identity (2.2) is usually called the
generalized Jacobi identity, or simply the Jacobi identity.

A derivation of an $n$-Lie algebra $A$
is a linear map $D$ of $A$ into itself satisfying
$$
 D([x_1, \cdots, x_n])=\sum_{i=1}^n[x_1, \cdots, D(x_i), \cdots, x_n]\eqno(2.3)
$$
for any $x_1, \cdots, x_n\in A$. Let $\mbox{Der}(A)$ be the set of
all derivations of $A$. Then $\mbox{Der}(A)$ is a Lie subalgebra of
the general linear Lie algebra $gl(A)$ and is called the derivation
algebra of $A$. The map  ad$(x_1, \cdots, x_{n-1})$: $A \rightarrow
A$, given by
$$\mbox{ad}(x_1, \cdots, x_{n-1})(x_n)=[x_1, \cdots, x_{n}], \mbox{ for } x_n\in A, $$
is referred to as a left multiplication defined by elements $x_1$,
$\cdots$, $x_{n-1} \in A$. It follows from identity (2.2), that
ad$(x_1, \cdots, x_{n-1})$ is a derivation. The set of all finite
linear combinations of left multiplications is an ideal of
$\mbox{Der}(A)$, which we denote by $\mbox{ad}(A)$. Every derivation
in $\mbox{ad}(A)$ is by definition an inner derivation.

If a subspace $B$ of an $n$-Lie algebra $A$ satisfying $[x_1,
\cdots, x_n]\in B$ for any $x_1, \cdots, x_n\in B$, then $B$ is
called a subalgebra of $A$. Let $A_{1}, A_{2}, \cdots, A_{n}$ be
subalgebras of an $n$-Lie algebra $A$.
 Denote by $[A_{1}, A_{2}, \cdots, A_{n}]$ the subspace of $A$ generated by all
 vectors $[x_{1}, \cdots, x_{n}]$, where $x_{i}\in A_{i}$ for $i=1, 2, \cdots, n$.
 The subalgebra $A^1 = [A, A, \cdots, A]$ is called the derived algebra of $A$. If $A^1=0,$
 then $A$ is called an abelian $n$-Lie algebra.

Let $H$ be an abelian subalgebra of $n$-Lie algebra $A$. Then $H$ is
by definition a Toral subalgebra of $A$, if $A$ is a complete
$H$-module, that is
$$ A=\oplus_{\alpha\in (H^{n-1})^{\ast}}A _{\alpha}  ~ ~\mbox{(direct sum as
vector spaces)},
$$
 where
 $$ A_{\alpha}=\{x \in
A~|~\mbox{ad}(h_1, \cdots, h_{n-1})(x)=\alpha(h_1, \cdots,
h_{n-1})(x),~\forall(h_{1}, h_{2}, \cdots, h_{n-1})\in H^{n-1}\}.$$

A Toral subalgebra $H$ is called maximal if there are no
 Toral subalgebras of $A$ properly containing $H.$ An ideal $I$ of an $n$-Lie algebra $A$ is a subspace of $A$ such
that $[I, A, \cdots, A]\subseteq I. $ If $ [I, I, A, \cdots, A]=0$,
 then $I$ is referred to as an abelian ideal. If $A^1\neq 0$ and $A$ has
 no ideals except $0$ and itself, then $A$ is by definition a simple
 $n$-Lie algebra. An $n$-Lie algebra $A$
  is said to be decomposable if there are nonzero ideals $I_1, I_2 $ such that
$$
    A = I_1\oplus I_2,
$$
then $[I_1, I_2, A, \cdots, A]=0 $. Otherwise, we say that $A$ is
indecomposable.
 Clearly if $A$ is a simple $n$-Lie algebra then $A$ is indecomposable.

The subset $Z(A)=\{ x\in A~|~ [x, y_1, \cdots, y_{n-1}]=0, ~\forall
~y_1, \cdots, y_{n-1}\in A\}$ is called the center of $A$. It is
clear that $Z(A)$ is an abelian ideal of $A$.

\vspace{2mm}\noindent{ \bf 3. Classification of $(n+2)$-dimensional
$n$-Lie algebras }

In this section, unless stated otherwise, we suppose that $F$ is an
algebraically closed field of characteristic $0$. Any brackets
of basis vectors not listed in the multiplication table of $n$-Lie
algebras are assumed to be zero.

First, we prove the  isomorphic criterion theorem for
$(n+2)$-dimensional $n$-Lie algebras over $F$.

We need some symbols for reducing our description. Suppose $[,
\cdots, ]_1$ and $[, \cdots, ]_2$ are two $n$-ary Lie products on
vector space $A$ such that $(A, [, \cdots, ]_1)$ and $( A, [,
\cdots, ]_2 )$ are $n$-Lie algebras. Let  $e_{1},$ $ e_{2},$ $
\cdots, $ $e_{n+2}$ be a basis of $A$. Set
$$
e_{i, j}=[e_{1}, \cdots, \hat{e}_{i}, \cdots, \hat{e}_{j}, \cdots,
e_{n+2}]_1=\sum\limits_{k=1}^{n+2}b^{k}_{i, j} e_{k}, ~b^{k}_{i,
j}\in F, 1\leq i < j \leq n+2, \eqno (3.1)
$$
then
$$(e_{1, 2}, e_{1, 3}, \cdots, e_{1, n+2}, e_{2, 3}, \cdots, e_{2, n+2}, \cdots, e_{n+1, n+2})=(e_{1},
e_{2}, \cdots, e_{n+2})B,$$ where $$B=\left(
      \begin{array}{ccccccc}
        b^{1}_{1, 2} & b^{1}_{1, 3} & \cdots & b^{1}_{1, n+2} & b^{1}_{2, 3} & \cdots & b^{1}_{n+1, n+2} \\
        b^{2}_{1, 2} & b^{2}_{1, 3} & \cdots & b^{2}_{1, n+2} & b^{2}_{2, 3} & \cdots & b^{2}_{n+1, n+2} \\
        \vdots & \vdots & \vdots & \vdots & \vdots & \vdots & \vdots \\
        b^{n+2}_{1, 2} & b^{n+2}_{1, 3} & \cdots & b^{n+2}_{1, n+2} & b^{n+2}_{2, 3} &
        \cdots & b^{n+2}_{n+1, n+2} \\
      \end{array}
    \right), ~b^{k}_{i,
j}\in F, 1\leq i < j \leq n+2.
$$

Then the multiplication of $(A,[, \cdots, ]_1) $ is determined by
the $((n+2)\times \frac{(n+1)(n+2)}{2})$ matrix $B$. And $B$ is
called
  the structure  matrix of $(A, [, \cdots, ]_1)$ with respect to
the basis $e_{1},$ $ e_{2},$ $ \cdots, $ $e_{n+2}$.

Similarly denote $\bar{B}$ is the structure  matrix of $(A, [,
\cdots, ]_2)$ with respect to the basis $e_{1},$ $ e_{2},$ $ \cdots,
$ $e_{n+2}$, that is
$$\bar{e}_{ij}=[e_{1}, \cdots, \hat{e}_{i}, \cdots, \hat{e}_{j},
\cdots, e_{n+2}]_2=\sum\limits_{k=1}^{n+2}\bar{b}^{k}_{i, j} e_{k},
~\bar{b}^{k}_{i, j}\in F, 1\leq i < j \leq n+2, \eqno(3.2) $$
$$
(\bar{e}_{1, 2}, \bar{e}_{1, 3}, \cdots, \bar{e}_{1, n+2},
\bar{e}_{2, 3},\cdots, \bar{e}_{2, n+2}, \cdots, \bar{e}_{n+1,
n+2})= (e_{1},  \cdots, e_{n+2})\bar{B}.
$$

\vspace{1mm}\noindent {\bf Theorem 3.1.} $N$-Lie algebras $(A, [,
\cdots,]_1)$ and $(A, [, \cdots,]_2)$ with products (3.1) and (3.2)
on an $(n+2)$-dimensional linear space $A$ are isomorphic if and
only if there exists a nonsingular $((n+2)\times(n+2))$ matrix
$T=(t_{i, j})$ such that
$$
B=T'^{-1}\bar{B}T_*,\eqno(3.3)
$$ where $T'$ is the transpose matrix of $T$, and $T_{*}=(T_{k, l}^{i, j})$ is an
$(\frac{(n+1)(n+2)}{2}\times \frac{(n+1)(n+2)}{2})$ matrix, and
$T^{i, j}_{k, l}\in F$ is the determinant defined by (3.5) below for
$1\leq i, j, k, l\leq n+2$.

\noindent{\bf Proof.} If $n$-Lie algebra $(A, [, \cdots, ]_1)$ is
isomorphic to $(A, [, \cdots, ]_2)$ under the isomorphism $\sigma$.
Let $e_1, \cdots, e_{n+2}$ be a basis of $A$, and structural
matrices are (3.1) and (3.2) with respect to $e_1, \cdots, e_{n+2}$
respectively, that is
$$ e_{i, j}=[e_{1}, \cdots, \hat{e}_{i},
\cdots, \hat{e}_{j}, \cdots,
e_{n+2}]_1=\sum\limits_{k=1}^{n+2}b^{k}_{i, j} e_{k},~B=(b_{i,
j}^k)_{(n+2)\times\frac{(n+2)\times (n+2)}{2}};
$$
and
$$ \bar{e}_{i, j}=[e_{1}, \cdots, \hat{e}_{i}, \cdots,
\hat{e}_{j}, \cdots,
e_{n+2}]_2=\sum\limits_{k=1}^{n+2}\bar{b}^{k}_{i, j} e_{k},
~~\bar{B}=(\bar{b}_{i, j}^k)_{(n+2)\times\frac{(n+2)\times
(n+2)}{2}}.
$$

Denote $e'_i=\sigma(e_i), ~1\leq i\leq n+2$ and the nonsingular
$((n+2)\times (n+2))$ matrix $T=(t_{ij})$ is the transition matrix
of $\sigma$ in the basis $e_{1}, $ $e_{2}, $ $\cdots, $ $e_{n+2},$
that is
$$
(\sigma(e_{1}),  \cdots, \sigma(e_{n+2}))=(e'_{1},  \cdots,
e'_{n+2})=(e_{1}, e_{2}, \cdots, e_{n+2})T.\eqno(3.4)$$

Then $$e_{k, l}'=[e'_{1}, \cdots, \hat{e'}_{k}, \cdots,
\hat{e'}_{l}, \cdots, e'_{n+2}]_2$$ \vspace{1mm}
$$=[\sum\limits_{m=1}^{n+2}t_{m, 1} e_{m}, \sum\limits_{m=1}^{n+2}t_{m, 2}
e_{m}, \cdots,~\sum\limits_{m=1}^{n+2}t_{m, k-1} e_{m},
\sum\limits_{m=1}^{n+2}t_{m, k+1} e_{m},
$$
$$
\cdots, \sum\limits_{m=1}^{n+2}t_{m, l-1} e_{m},
\sum\limits_{m=1}^{n+2}t_{m,l+1}e_{m}, \cdots,
\sum\limits_{m=1}^{n+2}t_{m, n+2} e_{m}]_2 $$
$$=T_{k, l}^{1, 2}\bar{e}_{1, 2}+T_{k, l}^{1, 3}\bar{e}_{1, 3}+ \cdots+T_{k, l}^{1, n+2}\bar{e}_{1, n+2}
+T_{k, l}^{2, 3}\bar{e}_{2, 3}+ \cdots+T_{k, l}^{n+1,
n+2}\bar{e}_{n+1, n+2},$$ where
$$T_{k, l}^{i, j}= \mbox{det}\left(
  \begin{array}{cccccccccc}
t_{1, 1}  & \cdots & t_{1, k-1}& t_{1, k+1}& \cdots & t_{1, l-1}& t_{1, l+1}&\cdots & t_{1, n+2} \\
t_{2, 1}  & \cdots & t_{2, k-1}& t_{2, k+1}& \cdots & t_{2, l-1}& t_{2, l+1}& \cdots & t_{2, n+2} \\
\vdots  & \vdots & \vdots & \vdots & \vdots & \vdots & \vdots & \vdots & \vdots\\
t_{i-1, 1}  & \cdots & t_{i-1, k-1}& t_{i-1, k+1}& \cdots & t_{i-1,
l-1}& t_{i-1, l+1}&\cdots
& t_{i-1, n+2} \\
t_{i+1, 1}  & \cdots & t_{i+1, k-1}& t_{i+1, k+1}& \cdots & t_{i+1,
l-1}& t_{i+1, l+1}&\cdots
& t_{i+1, n+2} \\
\vdots  & \vdots & \vdots&\vdots & \vdots & \vdots & \vdots & \vdots& \vdots \\
t_{j-1, 1}  & \cdots & t_{j-1, k-1}& t_{j-1, k+1}& \cdots & t_{j-1,
l-1}& t_{j-1, l+1}&\cdots
& t_{j-1, n+2} \\
t_{j+1, 1}  & \cdots & t_{j+1, k-1}& t_{j+1, k+1}& \cdots & t_{j+1,
l-1}& t_{j+1, l+1}&\cdots
& t_{j+1, n+2} \\
\vdots  & \vdots & \vdots&\vdots & \vdots & \vdots & \vdots & \vdots& \vdots \\
t_{n+1, 1}  &\cdots& t_{n+1, k-1}& t_{n+1, k+1}& \cdots &
t_{n+1, l-1}& t_{n+1, l+1}&\cdots & t_{n+1, n+2} \\
t_{n+2, 1}  &\cdots& t_{n+2, k-1}& t_{n+2, k+1}& \cdots & t_{n+2,
l-1}& t_{n+2, l+1}& \cdots
& t_{n+2, n+2} \\
\end{array}
\right) \eqno(3.5)
$$ \vspace{1mm}  $1\leq i< j\leq n+2,~1\leq k\neq l\leq
n+2$. Denote
$$T_{*}=\left(
                      \begin{array}{ccccccc}
T_{1, 2}^{1, 2} & T_{1, 3}^{1, 2} & \cdots & T_{1, n+2}^{1, 2} & T_{2, 3}^{1, 2} & \cdots & T_{n+1, n+2}^{1, 2} \\
 T_{1, 2}^{1, 3} & T_{1, 3}^{1, 3} & \cdots & T_{1, n+2}^{1, 3} & T_{2, 3}^{1, 3} & \cdots & T_{n+1, n+2}^{1, 3} \\
 \vdots & \vdots & \vdots & \vdots & \vdots & \vdots & \vdots \\
 T_{1, 2}^{n, n+2} & T_{1, 3}^{n, n+2} & \cdots & T_{1, n+2}^{n, n+2} & T_{2, 3}^{n, n+2} &
  \cdots & T_{n+1, n+2}^{n, n+2} \\
T_{1, 2}^{n+1, n+2} & T_{1, 3}^{n+1, n+2} & \cdots & T_{1,
n+2}^{n+1, n+2} & T_{2, 3}^{n+1, n+2} &
\cdots & T_{n+1, n+2}^{n+1, n+2} \\
 \end{array}
 \right),\eqno(3.6)
$$
then $T_{*}$ is a $(\frac{(n+1)(n+2)}{2}\times \frac{(n+1)(n+2)}{2}
)$ matrix, and
$$(e_{1, 2}', e_{1, 3}', \cdots, e_{1, n+2}', e_{2, 3}', \cdots, e_{n+1, n+2}')=
(\bar{e}_{1, 2}, \bar{e}_{1, 3}, \cdots, \bar{e}_{1, n+2},
\bar{e}_{2, 3}, \cdots, \bar{e}_{n+1, n+2})T_{*}. \eqno (3.7)$$

From identities (3.1) and (3.2) that
$$ (e_{1, 2}', e_{1, 3}',
\cdots, e_{1, n+2}', e_{2, 3}', \cdots, e_{n+1, n+2}')= (e_{1},
e_{2}, \cdots, e_{n+2})\bar{B}T_{*}.\eqno(3.8)
$$ Furthermore
$$
e_{k, l}'=[e'_{1}, \cdots, \hat{e}_{k}', \cdots, \hat{e}_{l}',
\cdots, e'_{n+2}]_2 =[\sigma(e_1), \cdots, \hat{\sigma(e_{k})},
\cdots, \hat{\sigma(e_{l})}, \cdots, \sigma(e_{n+2})]_2
$$
$$
=\sigma([e_{1}, \cdots, \hat{e}_{k}, \cdots, \hat{e}_{l}, \cdots,
e_{n+2}]_1) =\sigma(e_{kl})=\sum\limits_{i=1}^{n+2}b^{i}_{kl}
\sigma(e_{i})=\sum\limits_{s=1}^{n+2}(\sum\limits_{i=1}^{n+2}b^{i}_{kl})t_{si}e_s.
$$
Thus
$$
(e_{1, 2}', e_{1, 3}', \cdots, e_{1, n+2}', e_{2, 3}', \cdots,
e_{n+1, n+2}')= (e_{1}, e_{2}, \cdots, e_{n+2})T'B.\eqno(3.9)$$

It follows from (3.8) and (3.9) that
$$
T'B=\bar{B}T_*, ~\mbox{that is}~B= T'^{-1}\bar{B}T_*.
$$

 On the other hand, we take a linear transformation $\sigma$ of $A$,
 such that $\sigma(e_1, \cdots, e_{n+2})=(e_1, \cdots, e_{n+2})T.$
By similar discussions to the above we have  $\sigma$ is an $n$-Lie
isomorphism from $(A, [ , \cdots, ]_1)$ to $(A, [ , \cdots, ]_2).$~~
\hfill$\Box$

It is complex when we use Theorem 3.1 to judge the isomorphism of
two $(n+2)$-dimensional $n$-Lie algebras due to the massive
computations involved. But from (3.5) and (3.3), the computation is
orderly so it is easy to use computer.

Before giving the classification theorem, we need to classify the
$(n+1)$-dimensional $n$-Lie algebras first.

 \vspace{1mm} \noindent{\bf  Lemma 3.1.} Let $A$ be an
$(n+1)$-dimensional $n$-Lie algebra over
   $F$ and  $e_{1},$ $ e_{2},$
  $ \cdots,$ $ e_{n+1}$ be a basis of $A$ ($n\geq 3$). Then one and only one
  of the following possibilities holds up
to isomorphisms:

\vspace{2mm}\noindent $(a)$ ~If $\dim A^{1}=0$, then $A$ is an
abelian $n$-Lie algebra.

 \vspace{2mm}\noindent \noindent $(b)$ ~If $\dim A^{1}=1$ and let $A^{1} =
F e_{1}$, then in the case that $A^{1}\subseteq Z(A)$,
$$ (b_1) ~ [e_{2}, \cdots, e_{n+1}] = e_{1};\eqno(3.10)$$
\noindent in the case that $A^{1}$ is not contained in $Z(A)$,
$$ (b_2) ~ [e_{1}, \cdots, e_{n}]=e_{1}. \eqno(3.11)$$

\vspace{2mm}\noindent  $(c)$ ~If $\dim A^{1}=2$ and let $A^{1}= F
e_{1}+ F e_{2}$, then
$$\begin{array}{ll}
(c_{1}) \left\{\begin{array}{l}
{[} e_{2}, \cdots, e_{n+1}] = e_{1},\\
{[}e_{1}, e_{3}, \cdots, e_{n+1}] = e_{2};
\end{array}\right.
&
 (c_{2}) \left\{\begin{array}{l}
{[} e_{2}, \cdots, e_{n+1}] =\alpha e_{1}+ e_{2},\\
{[}e_{1}, e_{3}, \cdots, e_{n+1}] = e_{2};
\end{array}\right.
\end{array}
$$
$$
\begin{array}{ll}
 (c_{3}) \left\{\begin{array}{l}
{[}e_{1}, e_{3}, \cdots, e_{n+1}] = e_{1}, \\
{[}e_{2}, \cdots, e_{n+1}] = e_{2},
\end{array}\right.
\end{array}                          \eqno(3.12)$$
where $~ \alpha \in F$ and $ \alpha \neq 0.$

\vspace{2mm}\noindent  $(d)$ ~If  $\dim A^{1}=r$, $ 3\leq r\leq
n+1$, let $A^{1}= F e_{1}+ F e_{2}+ \cdots +Fe_{r}$. Then
$$
(d_r) ~[e_1, \cdots, \hat{e}_i, \cdots, e_{n+1}]=e_i, ~1\le i \le r,
\eqno(3.13)
$$
 where symbol
$\hat{e}_i$
 means that  $e_{i}$ is omitted.

\vspace{2mm}\noindent  {\bf Proof.} If $\dim A^1=1$ or $\dim A^1>2$,
the classification has been discussed by [3]. Now we study the case
$\dim A^1=2.$  Set $A^{1}=Fe_{n}+Fe_{n+1}$,
$e^{i}=(-1)^{n+1+i}{[}e_{1},\cdots,\hat{e_{i}},\cdots,e_{n+1}]=\beta_{ni}e_{n}+\beta_{n+1i}e_{n+1},1\leq
i\leq n-1.$ Then we have
$$\begin{array}{ll} (c)~ \left\{\begin{array}{l}
e^{1}=(-1)^{n+1+1}{[}e_{2}, \cdots, e_{n+1}] =\beta_{11}e_{1}+\beta_{21}e_{2},\\
e^{2}=(-1)^{n+1+2}{[}e_{1}, e_{3}, \cdots, e_{n+1}] = \beta_{12}e_{1}+\beta_{22}e_{2},\\
e^{i}=(-1)^{n+1+i}{[}e_{1}, e_{2}, e_{3}, \cdots, \hat{e_{i}},
\cdots, e_{n+1}] =\beta_{1i}e_{1}+\beta_{2i}e_{2},(3\leq i \leq
n+1).
\end{array}\right.
\end{array}$$
 and
$$\left|
\begin{array}{ccc}
\beta_{11} & \beta_{12}\\
\beta_{21}& \beta_{22}\\
\end{array}
\right|\neq0 ,\quad\quad
(\beta_{i2}-\beta_{2i})e^{1}+(\beta_{1i}-\beta_{i1})e^{2}+(\beta_{21}-\beta_{12})e^{i}=0.\eqno(*)$$
It follows that $e^{i}$ and $e^{j}$ are linearly dependent for
$i\neq j$ for $ i, j=3, \cdots, n+1$. And by $\dim A^1=2$ and $(*),$
we have  $e^{i}=0$ for $3\leq i\leq n+1$. Then (c) is reduced to
$$(1) ~\begin{array}{ll}
\left\{\begin{array}{l}
e^{1}=(-1)^{n+1+1}{[}e_{2}, \cdots, e_{n+1}] =a e_{1}+c e_{2},\\
e^{2}=(-1)^{n+1+2}{[}e_{1}, e_{3}, \cdots, e_{n+1}] = b e_{1}+d
e_{2}.
\end{array}\right.
\end{array}~ A=\left(
\begin{array}{ccc}
a & b\\
c& d\\
\end{array}
\right), ~\det A\neq0. $$

If $a \neq0 $, let $P=\left(
                 \begin{array}{cc}
                   \frac{1}{\sqrt{a}} & 0\\
                  \frac{ -c}{\sqrt{(ad-bc)a}} & \frac{\sqrt{a}}{\sqrt{ad-bc}} \\
                 \end{array}
               \right)$, then
$$(\det P^{-1}) PAP'=\left(
                 \begin{array}{cc}
                   \sqrt{ad-bc} & b-c \\
                   0 & \sqrt{ad-bc} \\
                 \end{array}
               \right).
$$
By Theorem 2 in [3], (1) is isomorphic to
$$ (1)' ~\begin{array}{ll}
\left\{\begin{array}{l}
e^{1}=(-1)^{n+1+1}{[}e_{2}, \cdots, e_{n+1}] =\sqrt{ad-bc} e_{1},\\
e^{2}=(-1)^{n+1+2}{[}e_{1},e_{3}, \cdots, e_{n+1}] =(b-c)
e_{1}+\sqrt{ad-bc} e_{2},
\end{array}\right.
\end{array}$$

In the case of $b-c= 0$, substituting $e_1$ and $e_{n+1}$ by $ie_1$
and $(-1)^{n+1+1}\frac{1}{\sqrt{ad-bc}}ie_{n+1}$ respectively, we
get (1) is isomorphic to
$$  \begin{array}{ll}
(c_1)~ \left\{\begin{array}{l}
{[}e_{2}, \cdots, e_{n+1}]=e_{1},\\
{[}e_{1}, e_{3}, \cdots, e_{n+1}]=e_{2}.
\end{array}\right.
\end{array}$$

In the case of  $b-c\neq0$, substituting $e_1$ and $e_{n+1}$ by
$e_{1}+\frac{(b-c)}{\sqrt{ad-bc}} e_{2}$ and
$\frac{e_{n+1}}{\sqrt{ad-bc}}$ we get (1) is isomorphic to
$$\begin{array}{ll}
 ~\left\{\begin{array}{l}
{[}e_{2}, \cdots, e_{n+1}]=(-1)^{n+1+1}e_{1}-(-1)^{n+1+1}\frac{(b-c)}{\sqrt{ad-bc}}  e_{2},\\
{[}e_{1},e_{3}, \cdots, e_{n+1}]=(-1)^{n+1+2}e_{2},
\end{array}\right.
\end{array}$$
\par
substituting $e_1$,$e_2$ and $e_{n+1}$ by $ie_{1}$
and$-\frac{\sqrt{ad-bc}}{b-c}ie_2$,
$(-1)^{n+1+1}\frac{\sqrt{ad-bc}}{b-c}e_{n+1}$ we get (1) is
isomorphic to

$$\begin{array}{ll}
(c_2) ~\left\{\begin{array}{l}
{[}e_{2}, \cdots, e_{n+1}]=\alpha e_{1}+e_{2},\\
{[}e_{1}, e_{3}, \cdots, e_{n+1}]= e_{2},
\end{array}\right.
\end{array}$$

where $\alpha=-\frac{ad-bc}{(b-c)^{2}}\neq 0.$

If $a=0$, then (1) is of the form
$$\begin{array}{ll}
\left\{\begin{array}{l}
e^{1}=(-1)^{n+1+1}{[}e_{2}, \cdots, e_{n+1}] =c e_{2},\\
e^{2}=(-1)^{n+1+2}{[}e_{1},e_{3}, \cdots, e_{n+1}] = b e_{1}+d
e_{2}.
\end{array}\right.
\end{array} ~\det \left(
\begin{array}{ccc}
0 & b\\
c & d\\
\end{array}
\right)\neq0. $$ In the cases of $d\neq0$ or $b+c\neq0$, by the
similar discussion to above, we have (1) is isomorphic to the case
$(c_1)$ or $(c_2)$. In the case of $d=b+c=0$, (1) is isomorphic to
$$\begin{array}{ll}
(c_3) ~\left\{\begin{array}{l}
{[}e_{1}, e_{3}, \cdots, e_{n+1}] =e_{1},\\
{[}e_{2}, \cdots, e_{n+1}] =e_{2}.
\end{array}\right.
\end{array}$$~~
\hfill$\Box$

\vspace{2mm}\noindent  {\bf Lemma 3.2.}[22] Let $A$ be a nonabelian
$(n+2)$-dimensional $n$-Lie algebra over $F$. If $\dim A^1\neq 3$,
then there exists a non-abelian subalgebra of codimension $1$
containing $A^1$.

\vspace{2mm}\noindent  {\bf Lemma 3.3.}[22] Let $A$ be an
$(n+2)$-dimensional $n$-Lie algebra over $F$. Then we have $\dim
A^1\leq n+1$.

\vspace{2mm} \noindent{\bf Theorem 3.2.} Let $A$ be an
$(n+2)$-dimensional $n$-Lie algebra over $F$ with a basis $e_1,$
$\cdots,$ $ e_{n+2}$. Then one and only one of the following
possibilities holds up to isomorphisms:

\vspace{2mm}\noindent $(a)$~ If $\dim A^{1}=0$, then $A$ is an
abelian  $n$-Lie algebra.

\vspace{2mm}\noindent  $(b)$~  If $\dim A^{1}=1$, let $A^1=Fe_1$.
Then we have

\vspace{2mm}\noindent $(b^{1})$~ in the case that $A^{1}\subseteq
Z(A)$, ~${[}e_{2}, \cdots, e_{n+1}]=e_{1};$

\vspace{2mm}\noindent $(b^{2})$~ in the case that $A^{1}$ is not
contained in $Z(A)$, ~$[e_{1}, \cdots, e_{n}]=e_{1}.$

\vspace{2mm}\noindent $(c)$~ If $\dim A^{1}=2$, let $A^{1}= F e_{1}+
F e_{2}$. Then we have

\vspace{2mm}\noindent $\begin{array}{ll} (c^{1})~
\left\{\begin{array}{l}
{[}e_{2}, \cdots, e_{n+1}] = e_{1},\\
{[}e_{3}, \cdots, e_{n+2}] = e_{2};
\end{array}\right.
~~~~~ (c^{2})~ \left\{\begin{array}{l}
{[}e_{2}, \cdots, e_{n+1}] =  e_{1}, \\
{[}e_{2}, e_{4}, \cdots, e_{n+2}] = e_{2},\\
{[}e_{1}, e_{4}, \cdots, e_{n+2}] = e_{1};
\end{array}\right.
\end{array}
$

\vspace{2mm}\noindent $\begin{array}{ll} (c^{3})~
\left\{\begin{array}{l}
 {[}e_{2}, \cdots, e_{n+1}] = e_{1},\\
{[}e_{1}, e_{3}, \cdots, e_{n+1}] = e_{2};
\end{array}\right.
 ~(c^{4})~ \left\{\begin{array}{l}
{[}e_{2}, \cdots, e_{n+1}] = e_{1},\\
{[}e_{1}, e_{3}, \cdots, e_{n+1}] = e_{2}, \\
{[}e_{2}, e_{4}, \cdots, e_{n+2}] = e_{2},\\
{[}e_{1}, e_{4}, \cdots, e_{n+2}] = e_{1};
\end{array}\right.
\end{array}
$

\vspace{2mm}\noindent $\begin{array}{ll} (c^{5})~
\left\{\begin{array}{l}
{[}e_{2}, \cdots, e_{n+1}] =\alpha e_{1}+ e_{2},\\
{[}e_{1}, e_{3}, \cdots, e_{n+1}] = e_{2};
\end{array}\right.  ~ (c^{6})~ \left\{\begin{array}{l}
{[}e_{2}, \cdots, e_{n+1}] =\alpha e_{1}+e_{2},\\
{[}e_{1}, e_{3}, \cdots, e_{n+1}] = e_{2}, \\
{[}e_{2}, e_{4}, \cdots, e_{n+2}] = e_{2},\\
{[}e_{1}, e_{4}, \cdots, e_{n+2}] = e_{1};
\end{array}\right.
\end{array}$

\vspace{2mm}\noindent $\begin{array}{ll}
(c^{7})\left\{\begin{array}{l}
{[}e_{1}, e_{3}, \cdots, e_{n+1}] = e_{1},\\
{[}e_{2}, e_{3}, \cdots, e_{n+1}] = e_{2};\\
\end{array}\right.
\end{array}$

\vspace{2mm}\noindent where $\alpha\in F$, and $\alpha\neq 0$.

\vspace{2mm}\noindent $(d)$~ If $\dim A^{1}=3$, let $A^{1}= F e_{1}+
F e_{2}+ Fe_{3}$. Then we have

\vspace{2mm}\noindent $\begin{array}{ll} (d^{1})~
\left\{\begin{array}{l}
{[}e_{2}, \cdots, e_{n+1}] = e_{1},\\
{[}e_{2},e_{4},  \cdots, e_{n+2}]=-e_{2},\\
{[}e_{3}, \cdots, e_{n+2}] =e_{3};
\end{array}\right.
  (d^{2})~ \left\{\begin{array}{l}
{[}e_{2}, \cdots, e_{n+1}]=e_{1}, \\
{[}e_{3}, \cdots, e_{n+2}]=e_{3}+\alpha e_{2}, \\
{[}e_{2}, e_{4}, \cdots, e_{n+2}]=e_{3}, \\
{[}e_{1}, e_{4}, \cdots, e_{n+2}]=e_{1};
\end{array}\right.
\end{array}$

\vspace{2mm}\noindent $\begin{array}{ll} (d^{3})~
\left\{\begin{array}{l}
{[}e_{2}, \cdots, e_{n+1}]=e_{1}, \\
{[}e_{3}, e_{4}, \cdots, e_{n+2}]=e_{3}, \\
{[}e_{2}, e_{4}, \cdots, e_{n+2}]=e_{2}, \\
{[}e_{1}, e_{4}, \cdots, e_{n+2}]=2e_{1};
\end{array}\right.
(d^{4})~ \left\{\begin{array}{l}
{[}e_{2}, \cdots, e_{n+1}] = e_{1},\\
{[}e_{1}, e_{3}, \cdots, e_{n+1}] = e_{2},\\
{[}e_{1}, e_{2}, e_{4}, \cdots, e_{n+1}] = e_{3};
\end{array}\right.
\end{array}$

\vspace{2mm}\noindent $\begin{array}{ll}
 (d^{5})~ \left\{\begin{array}{l}
{[}e_{1}, e_{4}, \cdots, e_{n+2}]=e_{1},\\
{[}e_{2}, e_{4}, \cdots, e_{n+2}]= e_{3},\\
{[}e_{3}, e_{4}, \cdots, e_{n+2}]= \beta e_2+(1+\beta)e_{3}, ~
\beta\in F, \beta\neq 0, 1;
\end{array}\right.
\end{array}$

\vspace{2mm}\noindent $ \begin{array}{l} (d^{6})~
\left\{\begin{array}{l}
{[}e_{1}, e_{4}, \cdots, e_{n+2}]=e_{1},\\
{[}e_{2}, e_{4}, \cdots, e_{n+2}]=e_{2},\\
{[}e_{3}, e_{4}, \cdots, e_{n+2}]=e_{3};
\end{array}\right.
\end{array}
$

\vspace{2mm}\noindent  $\begin{array}{l} (d^{7})~
\left\{\begin{array}{l}
{[}e_{1}, e_{4}, \cdots, e_{n+2}]=e_{2},\\
{[}e_{2}, e_{4}, \cdots, e_{n+2}]= e_{3},\\
{[}e_{3}, e_{4}, \cdots, e_{n+2}]= se_1+te_2+ue_{3}, ~s, t, u\in
F,~s\neq 0.
\end{array}\right.
\end{array}
$

\vspace{2mm}\noindent And $n$-Lie algebras corresponding to the case
$(d^{7})$ with coefficients  $s, t, u$ and $s', t', u'$ are
isomorphic  if and only if there exists a nonzero element $r\in F$
such that
$$
s=r^3 s', ~t=r^2 t', ~u=ru', ~s, s', t, t', u, u' \in F.
$$

\vspace{2mm}\noindent $(r)$ ~ If $\dim A^{1}=r, 4\leq r\leq n+1$,
let $A^{1}= F e_{1}+ \ldots+ F e_{r}$. Then we have

\vspace{2mm}\noindent $\begin{array}{ll} (r^{1})~
\left\{\begin{array}{l}
{[}e_{2}, \cdots, e_{n+1}] = e_{1},\\
{[}e_{3}, \cdots, e_{n+2}] = e_{2},\\
 \ldots  \quad  \ldots \quad \ldots \quad \ldots ,\\
{[}e_{2}, \cdots, \hat{e_i}, \cdots, e_{r}, \cdots, e_{n+2}]=e_{i},\\
 \ldots  \quad  \ldots \quad \ldots \quad \ldots ,\\
{[}e_{2}, \cdots, e_{r-1}, e_{r+1}, \cdots, e_{n+2}] = e_{r};
\end{array}\right.
(r^{2})~ \left\{\begin{array}{l}
{[}e_{2}, \cdots, e_{n+1}] = e_{1},\\
 \ldots  \quad  \ldots \quad \ldots \quad \ldots ,\\
{[}e_{1}, \cdots, \hat{e_i}, \cdots, e_{r}, \cdots, e_{n+1}] = e_{i},\\
 \ldots  \quad  \ldots \quad \ldots \quad \ldots , \\
 {[}e_{1}, \cdots, e_{r-1}, e_{r+1}, \cdots, e_{n+1}] = e_{r}.
\end{array}\right.
\end{array}$

\vspace{2mm}\noindent  {\bf Proof.} 1. Case $(a)$ is trivial.

2. Case $(b)$. Suppose $A^1=F e_1$.
 Then from Lemma 3.1, Lemma 3.2 and
 Lemma 3.3, the multiplication table of $A$ in the basis $e_1, \cdots, e_{n+2}$
has the following  possibilities

\vspace{2mm}\noindent $\begin{array}{ll} (1)~
\left\{\begin{array}{l}
{[}e_{2}, \cdots, e_{n+1}] = e_{1},\\
{[}e_{1}, \cdots, \hat{e_{i}}, \cdots, \hat{e_{j}}, \cdots, e_{n+1},
e_{n+2}] = b_{ij}e_{1};
\end{array}\right.
\end{array}$

\vspace{2mm}\noindent $\begin{array}{ll} (2)~
\left\{\begin{array}{l}
{[}e_{1}, \cdots, e_{n}] = e_{1},\\
{[}e_{1}, \cdots, \hat{e_{i}}, \cdots, \hat{e_{j}}, \cdots, e_{n+1},
e_{n+2}] = b_{ij}e_{1},
\end{array}\right.
\end{array}$

\vspace{2mm}\noindent where $b_{ij}\in F, 1\leq i< j\leq n+1$.

Firstly, substituting the first identity of (1) into its other
equations and using the Jacobi identities, we get

 \vspace{2mm}\noindent $
b_{ij}e_{1}={[}e_{1}, e_{2}, \cdots, \hat{e_{i}}, \cdots,
\hat{e_{j}}, \cdots, e_{n+1}, e_{n+2}]$

\vspace{2mm}\noindent $ =[{[}e_{2}, \cdots, e_{n+1}], e_{2}, \cdots,
\hat{e_{i}}, \cdots, \hat{e_{j}}, \cdots, e_{n+1}, e_{n+2}]$

\vspace{2mm}\noindent $=[e_{2}, \cdots, [e_{i}, e_{2}, \cdots,
\hat{e_{i}}, \cdots, \hat{e_{j}}, \cdots, e_{n+1}, e_{n+2}],
\cdots,e_{n+1}]$

\vspace{2mm}\noindent $+[e_{2}, \cdots, [e_{j}, e_{2}, \cdots,
\hat{e_{i}}, \cdots, \hat{e_{j}}, \cdots, e_{n+1}, e_{n+2}],
\cdots,e_{n+1}]$

\vspace{2mm}\noindent $=[e_{2}, \cdots, (-1)^{i-2}b_{1j}e_{1},
\cdots, e_{n+1}]+[e_{2}, \cdots, (-1)^{j-3}b_{1i}e_{1}, \cdots,
e_{n+1}]$

\vspace{2mm}\noindent $=b_{1j}[e_{1}, e_{2}, \cdots, \hat{e_{i}},
\cdots, e_{n+1}]+b_{1i}[e_{1}, e_{2}, \cdots, \hat{e_{j}}, \cdots,
\cdots e_{n+1}]=0,$
 $2\leq i< j\leq n+1$. Then (1) is in the form of

\vspace{2mm}\noindent $\begin{array}{ll}
 (1)'~ \left\{\begin{array}{l}
{[}e_{2}, \cdots, e_{n+1}] = e_{1},\\
{[}e_{2}, \cdots, \hat{e_{j}}, \cdots, e_{n+1}, e_{n+2}] =
b_{1j}e_{1}, ~ 2\leq j\leq n+1.
\end{array}\right.
\end{array}$

\vspace{2mm}\noindent Replacing $e_{n+2}$ by
$e_{n+2}-\sum\limits_{j=2}^{n+1}(-1)^{n+1-j}b_{1j}e_{j}$ in $(1)'$,
we get that (1) is isomorphic to

\vspace{2mm}\noindent $(b^{1}) ~{[}e_{2}, \cdots, e_{n+1}]=e_{1}. $

\vspace{2mm} By similar discussion we get that (2) is isomorphic to
$(b^2)$. And $(b^1)$ is not isomorphic to $(b^2)$ since $(b^1)$ has
a nonzero center.

\vspace{2mm} 3. If $\dim A^{1}=2$, suppose $A^{1}=Fe_{1}+Fe_{2}$. By
Lemma 3.1, Lemma 3.2 and Lemma 3.3,  the multiplication table in the
basis $e_{1}, \cdots, e_{n+2}$ has the following possibilities

\vspace{2mm}\noindent $\begin{array}{ll} (1)~
\left\{\begin{array}{l}
{[}e_{2}, \cdots, e_{n+1}] = e_{1},\\
{[}e_{1}, \cdots, \hat{e_{i}}, \cdots, \hat{e_{j}}, \cdots, e_{n+1},
e_{n+2}] = b_{ij}^{1}e_{1}+b_{ij}^{2}e_{2};
\end{array}\right.
\end{array}$

\vspace{2mm}\noindent $\begin{array}{ll} (2)~
\left\{\begin{array}{l}
{[}e_{1}, \cdots, e_{n}] = e_{1},\\
{[}e_{1}, \cdots, \hat{e_{i}}, \cdots, \hat{e_{j}}, \cdots, e_{n+1},
e_{n+2}] =  b_{ij}^{1}e_{1}+b_{ij}^{2}e_{2};
\end{array}\right.
\end{array}$

\vspace{2mm}\noindent $\begin{array}{ll} (3)~
\left\{\begin{array}{l}
{[}e_{2}, \cdots, e_{n+1}] = e_{1},\\
{[}e_{1}, e_{3}, \cdots, e_{n+1}] = e_{2},\\
{[}e_{1}, \cdots, \hat{e_{i}}, \cdots, \hat{e_{j}}, \cdots, e_{n+1},
e_{n+2}] = b_{ij}^{1}e_{1}+b_{ij}^{2}e_{2};
\end{array}\right.
\end{array}$

\vspace{2mm}\noindent $\begin{array}{ll} (4)~
\left\{\begin{array}{l}
{[}e_{2}, \cdots, e_{n+1}] = \alpha e_{1}+e_{2},\\
{[}e_{1}, e_{3}, \cdots, e_{n+1}] = e_{2},\\
{[}e_{1}, \cdots, \hat{e_{i}}, \cdots, \hat{e_{j}}, \cdots, e_{n+1},
e_{n+2}] = b_{ij}^{1}e_{1}+b_{ij}^{2}e_{2},
\end{array}\right.
\end{array}$

\vspace{2mm}\noindent $\begin{array}{ll} (5)~
\left\{\begin{array}{l}
{[}e_{1}, e_{3}, \cdots, e_{n+1}] = e_{1},\\
{[}e_{2}, e_{3}, \cdots, e_{n+1}] = e_{2},\\
{[}e_{1}, \cdots, \hat{e_{i}}, \cdots, \hat{e_{j}}, \cdots, e_{n+1},
e_{n+2}] = b_{ij}^{1}e_{1}+b_{ij}^{2}e_{2};
\end{array}\right.
\end{array}$

\vspace{2mm}\noindent where $b_{ij}\in F, ~1\leq i< j\leq n+1$.

Firstly imposing the Jacobi identities on  (1) we get

\vspace{2mm}\noindent $b_{ij}^{1}e_{1}+b_{ij}^{2}e_{2}=[e_{1},
e_{2}, e_{3}, \cdots, \hat{e_{i}}, \cdots, \hat{e_{j}}, \cdots,
e_{n+1}, e_{n+2}]$

\vspace{2mm}\noindent $=[[e_{2}, \cdots, e_{n+1}], e_{2}, e_{3},
\cdots, \hat{e_{i}}, \cdots, \hat{e_{j}}, \cdots, e_{n+1}, e_{n+2}]$

\vspace{2mm}\noindent $=[e_{2}, \cdots, [e_{i}, e_{2}, e_{3},
\cdots, \hat{e_{i}}, \cdots, \hat{e_{j}}, \cdots, e_{n+1}, e_{n+2}],
\cdots, e_{n+1}]$

 \vspace{2mm}\noindent $+[e_{2}, \cdots,
[e_{j}, e_{2}, e_{3}, \cdots, \hat{e_{i}}, \cdots, \hat{e_{j}},
\cdots, e_{n+1}, e_{n+2}], \cdots,  e_{n+1}]$

\vspace{2mm}\noindent $=[e_{2}, \cdots,
(-1)^{i-2}(b_{1j}^{1}e_{1}+b_{1j}^{2}e_{2}), \cdots, e_{n+1}] +
[e_{2}, \cdots, (-1)^{j-3}(b_{1i}^{1}e_{1}+b_{1i}^{2}e_{2}), \cdots,
e_{n+1}]$
\par$=0,$ ~for  $3\leq i < j\leq n+1.$

\vspace{2mm}\noindent When $i=2$ and $3\leq j\leq n+1$,

\vspace{2mm}\noindent $b_{2j}^{1}e_{1}+b_{2j}^{2}e_{2}$ $=[e_{1},
e_{3}, \cdots, \hat{e_{j}}, \cdots, e_{n+1}, e_{n+2}] $

\vspace{2mm}\noindent $=[[e_{2}, \cdots, e_{n+1}], e_{3}, \cdots,
\hat{e_{j}}, \cdots, e_{n+1}, e_{n+2}]$

\vspace{2mm}\noindent $=[b_{1j}^{1}e_{1}+b_{1j}^{2}e_{2}, e_{3},
\cdots, e_{n+1}]+[e_{2}, \cdots,
(-1)^{j-3}(b_{12}^{1}e_{1}+b_{12}^{2}e_{2}), \cdots, e_{n+1}]$
$=b_{1j}^{2}e_{1}.$

\vspace{2mm}\noindent And again replacing $e_{n+2}$ by
$e_{n+2}-\sum\limits_{j=2}^{n+1}(-1)^{n+1-j}b_{1j}^{1}e_{j}$, we get

\vspace{2mm}\noindent $\begin{array}{ll}
 (1)' ~\left\{\begin{array}{l}
{[}e_{2}, \cdots, e_{n+1}] = e_{1},\\
{[}e_{3}, \cdots, e_{n+2}] = b_{12}^{2}e_{2},\\
{[}e_{2}, e_{3}, \cdots, \hat{e_{j}} \cdots, e_{n+1}, e_{n+2}] = b_{1j}^{2}e_{2},\\
{[}e_{1}, e_{3}, \cdots, \hat{e_{j}} \cdots, e_{n+1}, e_{n+2}] =
b_{1j}^{2}e_{1},
\end{array}\right.
\end{array}~ 3\leq j \leq n+1.$

\vspace{2mm} If $b_{12}^{2}\neq0, ~b_{1j}^{2}=0, 3\leq j\leq n+1, $
substituting $\frac{e_{n+2}}{b_{12}^{2}}$ for $e_{n+2}$ in $(1)'$,
we get (1) is isomorphic to

\vspace{2mm}\noindent $\begin{array}{ll} (c^{1})~
\left\{\begin{array}{l}
{[}e_{2}, \cdots, e_{n+1}] = e_{1},\\
{[}e_{3}, \cdots, e_{n+2}] = e_{2}.
\end{array}\right.
\end{array}$

\vspace{2mm}\noindent If there exists $j$ such that
$b_{1j}^{2}\neq0, 3\leq j\leq n+1$, then we might as well suppose
$b_{13}^{2}\neq0$. Substituting
$e_{3}+\sum\limits_{j=4}^{n+1}(-1)^{j-3}\frac{b_{1j}^{2}}{b_{13}^{2}}e_{j}-\frac{b_{12}^{2}}{b_{13}^{2}}e_{2}$
for $e_{3}$ and $\frac{e_{n+2}}{b_{13}^{2}}$ for $e_{n+2}$ in
$(1)'$, we get

\vspace{2mm}\noindent $\begin{array}{ll} (c^{2})~
\left\{\begin{array}{l}
{[}e_{2}, \cdots, e_{n+1}] = e_{1},\\
{[}e_{2}, e_{4}, \cdots, e_{n+2}] = e_{2},\\
{[}e_{1}, e_{4}, \cdots, e_{n+2}] = e_{1}.
\end{array}\right.
\end{array}$

\vspace{2mm} Secondly, substituting
$e_{n+2}-\sum\limits_{i=1}^{n}(-1)^{n-i}b_{in+1}^{1}e_{i}$ for
$e_{n+2}$ in (2), we get

\vspace{2mm}\noindent  $\begin{array}{ll}
 \left\{\begin{array}{l}
{[}e_{1}, \cdots, e_{n}] = e_{1},\\
{[}e_{1}, \cdots, \hat{e_{i}}, \cdots, \hat{e_{j}}, \cdots, e_{n},
e_{n+1}, e_{n+2}] = b_{ij}^{1}e_{1}+b_{ij}^{2}e_{2},
~1\leq i< j\leq n,\\
{[}e_{1}, \cdots, \hat{e_{i}}, \cdots, e_{n}, e_{n+2}] =
b_{in+1}^{2}e_{2}, ~1\leq i\leq n.
\end{array}\right.
\end{array}$

\vspace{2mm}\noindent Since $b_{2n+1}^{2}e_{2}=[e_{1}, e_{3} \cdots,
e_{n}, e_{n+2}]=[[e_{1}, \cdots, e_{n}],e_{3}, \cdots, e_{n},
e_{n+2}]$

\vspace{2mm}\noindent $=[b_{2n+1}^{2}e_{2},e_{2}, \cdots,
e_{n}]+[e_{1}, b_{1n+1}^{2}e_{2}, e_{3}, \cdots,
e_{n}]=b_{1n+1}^{2}e_{1},$ and

\vspace{2mm}\noindent $b_{in+1}^{2}e_{2}=[e_{1}, e_{2}, e_{3},
\cdots, \hat{e_{i}}, \cdots, e_{n}, e_{n+2}]=[[e_{1}, \cdots,
e_{n}], e_{2}, e_{3}, \cdots, \hat{e_{i}}, \cdots, e_{n}, e_{n+2}]$

\vspace{2mm}\noindent $=[b_{in+1}^{2}e_{2}, e_{2}, \cdots,
e_{n}]+[e_{1}, e_{2}, e_{3}, \cdots, (-1)^{i-2}b_{1n+1}^{2}e_{2},
\cdots, e_{n}]=0,$

 \vspace{2mm}\noindent we have $b_{2n+1}^{2}=b_{1n+1}^{2}=0$ and  $b_{in+1}^{2}=0, ~3\leq i\leq n.$
Then  (2) is isomorphic to

\vspace{2mm}\noindent  $\begin{array}{ll}
 \left\{\begin{array}{l}
{[}e_{1}, \cdots, e_{n}] = e_{1},\\
{[}e_{1}, \cdots, \hat{e_{i}}, \cdots, \hat{e_{j}}, \cdots, e_{n+1},
e_{n+2}] = b_{ij}^{1}e_{1}+b_{ij}^{2}e_{2}, ~1\leq i< j\leq n.
\end{array}\right.
\end{array}$

\vspace{2mm} When $i=1, ~2\leq j\leq n$, since

\vspace{2mm}\noindent $0=[[e_{1}, e_{3}, \cdots, e_{n}, e_{n+2}],
e_{2}, \cdots, \hat{e_{j}}, \cdots, e_{n}, e_{n+1}]$

\vspace{2mm}\noindent $=[e_{1}, e_{3}, \cdots, e_{n}, [e_{n+2},
e_{2}, \cdots, \hat{e_{j}}, \cdots, e_{n}, e_{n+1}]]$

\vspace{2mm}\noindent $=(-1)^{2n-3}b_{1j}^{2}e_{1},$
 we have $b_{1j}^{2}=0, ~2\leq j\leq n.$

\vspace{2mm} If $i=2, ~3\leq j\leq n$, by

\vspace{2mm}\noindent $b_{2j}^{1}e_{1}+b_{2j}^{2}e_{2}=[e_{1},
e_{3}, \cdots, \hat{e_{j}}, \cdots, e_{n}, e_{n+1}, e_{n+2}]$

\vspace{2mm}\noindent $=[[e_{1}, \cdots, e_{n}], e_{3}, \cdots,
\hat{e_{j}}, \cdots, e_{n}, e_{n+1}, e_{n+2}]$

\vspace{2mm}\noindent $=[b_{2j}^{1}e_{1}+b_{2j}^{2}e_{2}, e_{2},
\cdots, e_{n}] +[e_{1}, b_{1j}^{1}e_{1} +b_{1j}^{2}e_{2}, e_{3},
\cdots, e_{n}]$

\vspace{2mm}\noindent $+[e_{1}, e_{2}, e_{3}, \cdots, [e_{j}, e_{3},
\cdots, \hat{e_{j}}, \cdots, e_{n}, e_{n+1}, e_{n+2}], \cdots, e_{n}
]$

\vspace{2mm}\noindent
$=b_{2j}^{1}e_{1}+b_{1j}^{2}e_{1}=b_{2j}^{1}e_{1},$ \\we obtain $
b_{2j}^{2}=0, ~3\leq j\leq n$. If ~$3\leq i<j\leq n$, by

\vspace{2mm}\noindent $b_{ij}^{1}e_{1}+b_{ij}^{2}e_{2}=[e_{1},
e_{2}, \cdots, \hat{e_{i}}, \cdots, \hat{e_{j}}, \cdots, e_{n+1},
e_{n+2}]$

\vspace{2mm}\noindent $=[[e_{1}, \cdots, e_{n}], e_{2}, \cdots,
\hat{e_{i}}, \cdots, \hat{e_{j}}, \cdots, e_{n+1}, e_{n+2}]$

\vspace{2mm}\noindent $=[b_{ij}^{1}e_{1}+b_{ij}^{2}e_{2}, e_{2},
\cdots, e_{n}] +[e_{1}, e_{2}, \cdots, [e_{i}, e_{2}, e_{3}, \cdots,
\hat{e_{i}}, \cdots, \hat{e_{j}}, \cdots, e_{n+1}, e_{n+2}], \cdots,
e_{n}]$

\vspace{2mm}\noindent $+[e_{1}, e_{2}, \cdots, [e_{j}, e_{2}, e_{3},
\cdots, \hat{e_{i}}, \cdots, \hat{e_{j}}, e_{n+1}, e_{n+2}], \cdots,
e_{n}]=b_{ij}^{1}e_{1}, $

\vspace{2mm}\noindent we get $b_{ij}^{2}=0, 3\leq i<j\leq n.$ Then
(2)is isomorphic to

\vspace{2mm}\noindent $\begin{array}{ll}
 \left\{\begin{array}{l}
{[}e_{1}, \cdots, e_{n}] = e_{1},\\
{[}e_{1}, \cdots, \hat{e_{i}}, \cdots, \hat{e_{j}}, \cdots, e_{n+1},
e_{n+2}] =  b_{ij}^{1}e_{1}, ~1\leq i< j\leq n.
\end{array}\right.
\end{array}$

\vspace{2mm}\noindent This contradicts $\dim A^{1}=2$. Therefore,
table (2) is not realized.

Thirdly we study the case $(3)$. For $i=1,$ ~$3\leq j\leq n+1$,
since

\vspace{2mm}\noindent $b_{1j}^{1}e_{1}+b_{1j}^{2}e_{2}=[e_{2},
\cdots, \hat{e_{j}}, \cdots, e_{n+2}]=[[e_{1}, e_{3}, \cdots,
e_{n+1}], \cdots, \hat{e_{j}}, \cdots, e_{n+2}]$
$=b_{2j}^{1}e_{2}+b_{2j}^{2}e_{1},$ ~we have $b_{2j}^{1}=b_{1j}^{2},
b_{2j}^{2}=b_{1j}^{1},~ 3\leq j\leq n+1.$ For $3\leq i< j\leq n+1$,
from

\vspace{2mm}\noindent $b_{ij}^{1}e_{1}+b_{ij}^{2}e_{2}=[e_{1},
e_{2}, e_{3}, \cdots, \hat{e_{i}}, \cdots, \hat{e_{j}}, \cdots,
e_{n+1}, e_{n+2}]$

\vspace{2mm}\noindent $=[[e_{2}, \cdots, e_{n+1}], e_{2}, e_{3},
\cdots, \hat{e_{i}}, \cdots, \hat{e_{j}}, \cdots, e_{n+1}, e_{n+2}]$

\vspace{2mm}\noindent  $=[e_{2}, \cdots, [e_{i}, e_{2}, e_{3},
\cdots, \hat{e_{i}}, \cdots, \hat{e_{j}}, \cdots, e_{n+1}, e_{n+2}],
\cdots, e_{n+1}]$

\vspace{2mm}\noindent  $+[e_{2}, \cdots, [e_{j}, e_{2}, e_{3},
\cdots, \hat{e_{i}}, \cdots, \hat{e_{j}}, \cdots, e_{n+1}, e_{n+2}],
\cdots, e_{n+1}]=0,$

\vspace{2mm}\noindent  we have $ b_{ij}^{1}=b_{ij}^{2}=0, ~3\leq i<
j\leq n+1.$  Again substituting
$e_{n+2}+\sum\limits_{j=2}^{n+1}(-1)^{n+2-j}b_{1j}^{1}e_{j}+(-1)^{n}b_{12}^{2}e_{1}$
for $e_{n+2}$, (3) is isomorphic to

\vspace{2mm}\noindent $\begin{array}{ll}
 (3)' ~\left\{\begin{array}{l}
{[}e_{2}, \cdots, e_{n+1}] = e_{1},\\
{[}e_{1}, e_{3}, \cdots, e_{n+1}] = e_{2},\\
{[}e_{2}, e_{3}, \cdots, \hat{e_{j}}, \cdots, e_{n+1}, e_{n+2}] = b_{1j}^{2}e_{2}, \\
{[}e_{1}, e_{3}, \cdots, \hat{e_{j}}, \cdots, e_{n+1}, e_{n+2}] =
b_{1j}^{2}e_{1},
\end{array}\right.
\end{array} ~ ~3\leq j\leq n+1.$

\vspace{2mm}  If $b_{1j}^{2}=0, ~3\leq j\leq n+1$, then (3) is
isomorphic to

\vspace{2mm}\noindent $\begin{array}{ll} (c^{3})~
\left\{\begin{array}{l}
{[}e_{2}, \cdots, e_{n+1}] = e_{1},\\
{[}e_{1}, e_{3}, \cdots, e_{n+1}] = e_{2}.
\end{array}\right.
\end{array}$

\vspace{2mm} If there exists $b_{1j}^{2}\neq 0$ for $3\leq j\leq
n+1$, then we might as well suppose  $b_{13}^{2}\neq0$. Replacing
$e_{3}$ and $e_{n+2}$ by
$e_{3}+\sum\limits_{j=4}^{n+1}(-1)^{j-3}\frac{b_{1j}^{2}}{b_{13}^{2}}e_{j}$
and $\frac{e_{n+2}}{b_{13}^{2}}$ in $(3)'$ respectively, we get (3)
is of the form

\vspace{2mm}\noindent $\begin{array}{ll} (c^{4})~
\left\{\begin{array}{l}
{[}e_{2}, \cdots, e_{n+1}] = e_{1},\\
{[}e_{1}, e_{3}, \cdots, e_{n+1}] = e_{2},\\
{[}e_{2}, e_{4}, \cdots, e_{n+2}] = e_{2},\\
{[}e_{1}, e_{4}, \cdots, e_{n+2}] = e_{1}.
\end{array}\right.
\end{array}$

\vspace{2mm} Fourthly, we study the case (4). For $i=1, ~3\leq j\leq
n+1$, by $b_{1j}^{1}e_{1}+b_{1j}^{2}e_{2}=[e_{2}, \cdots,
\hat{e_{j}}, \cdots, e_{n+2}] =[[e_{1}, e_{3}, \cdots, e_{n+1}],
\cdots, \hat{e_{j}}, \cdots, e_{n+2}]$ $=b_{2j}^{1}
e_{2}+b_{2j}^{2}\alpha e_{1}+b_{2j}^{2} e_{2},$
 we have $b_{1j}^{1}=b_{2j}^{2}\alpha, b_{1j}^{2}=b_{2j}^{1}+b_{2j}^{2} , ~3\leq j\leq n+1.$

For $3\leq i< j\leq n+1$, we have $b_{ij}^{1}=b_{ij}^{2}=0, ~3\leq
i< j\leq n+1$ since

\vspace{2mm}\noindent  $b_{ij}^{1}e_{1}+b_{ij}^{2}e_{2}=[e_{1},
e_{2}, \cdots, \hat{e_{i}}, \cdots, \hat{e_{j}}, \cdots, e_{n+2}]$

\vspace{2mm}\noindent $=\frac{1}{\alpha}[[e_{2}, \cdots,
e_{n+1}]-e_{2}, e_{2}, \cdots, \hat{e_{i}}, \cdots, \hat{e_{j}},
\cdots, e_{n+2}]$

\vspace{2mm}\noindent $=\frac{1}{\alpha}[e_{2}, \cdots, [e_{i},
e_{2}, \cdots, \hat{e_{i}}, \cdots, \hat{e_{j}}, \cdots, e_{n+2}],
\cdots, e_{n+1}]$

\vspace{2mm}\noindent $+\frac{1}{\alpha}[e_{2}, \cdots, [e_{j},
e_{2}, \cdots, \hat{e_{i}}, \cdots, \hat{e_{j}}, \cdots, e_{n+2}],
\cdots, e_{n+1}]=0.$

\vspace{2mm}\noindent Then if substituting
$e_{n+2}+\sum\limits_{j=3}^{n+1}(-1)^{n+2-j}b_{2j}^{2}e_{j}+(-1)^{n}b_{12}^{2}e_{1}$
for $e_{n+2}$ in (4), we get

\vspace{2mm}\noindent  $\begin{array}{ll}
 (4)' ~\left\{\begin{array}{l}
{[}e_{2}, \cdots, e_{n+1}] =\alpha e_{1}+ e_{2},\\
{[}e_{1}, e_{3}, \cdots, e_{n+1}] = e_{2},\\
{[}e_{3}, \cdots, e_{n+2}] = b_{12}^{1}e_{1},\\
{[}e_{2}, e_{3}, \cdots, \hat{e_{j}}, \cdots, e_{n+1}, e_{n+2}] = b_{2j}^{1}e_{2} ,\\
{[}e_{1}, e_{3}, \cdots, \hat{e_{j}}, \cdots, e_{n+1},
e_{n+2}]=b_{2j}^{1}e_{1},
\end{array}\right.
\end{array} ~ 3\leq j\leq n+1.$

\vspace{2mm}\noindent  If $b_{12}^{1}=b_{2j}^{1}=0, 3\leq j\leq
n+1$, (4) is isomorphic to

\vspace{2mm}\noindent $\begin{array}{ll} (c^{5})~
\left\{\begin{array}{l}
{[}e_{2}, \cdots, e_{n+1}] =\alpha  e_{1}+e_{2},\\
{[}e_{1}, e_{3}, \cdots, e_{n+1}] = e_{2}.
\end{array}\right.
\end{array}$

\vspace{2mm} If $b_{12}^{1}\neq 0, b_{2j}^{1}=0, ~3\leq j\leq n+1$,
replacing $e_{n+2}$ by
$(-1)^{n}\frac{\alpha}{b_{12}^{1}}e_{n+2}-e_{1}+e_{2}$ in $(4)'$, we
get $(c^5)$.

If there exists $b_{2j}^{1}\neq 0,$ for some $3\leq j\leq n+1$. We
might as well suppose $b_{23}^{1}\neq 0$. Substituting
$e_{3}+\sum\limits_{j=4}^{n+1}(-1)^{j-3}\frac{b_{2j}^{1}}{b_{23}^{1}}e_{j}-\frac{b_{12}^{1}}{b_{23}^{1}}e_{1}$
and $\frac{e_{n+2}}{b_{23}^{1}}$ for $e_{3}$ and  $e_{n+2}$ in
$(4)'$ respectively, we get

\vspace{2mm}\noindent $\begin{array}{ll} (c^{6})~
\left\{\begin{array}{l}
{[}e_{2}, \cdots, e_{n+1}] =\alpha e_{1}+ e_{2},\\
{[}e_{1}, e_{3}, \cdots, e_{n+1}] = e_{2},\\
{[}e_{2}, e_{4}, \cdots, e_{n+2}] = e_{2},\\
{[}e_{1}, e_{4}, \cdots, e_{n+2}] = e_{1}.
\end{array}\right.
\end{array}$

\vspace{2mm} Lastly, we study the case (5). For $3\leq i< j\leq
n+1$, we have $b_{ij}^{1}=b_{ij}^{2}=0$ since

 \vspace{2mm}\noindent

 \quad $ [e_{1}, e_{2}, \cdots, \hat{e_{i}}, \cdots, \hat{e_{j}}, \cdots,
e_{n+2}].$

$=[{[}e_{1}, e_{2}, \cdots, \hat{e_{i}}, \cdots, \hat{e_{j}},
\cdots, e_{n+2}], e_{3}, \cdots, e_{n+1}]$

\vspace{2mm}\noindent$=[e_{1}, e_{2}, \cdots, \hat{e_{i}}, \cdots,
\hat{e_{j}}, \cdots, e_{n+2}] +[e_{1}, e_{2}, \cdots, \hat{e_{i}},
\cdots, \hat{e_{j}}, \cdots, e_{n+2}]$

Then if substituting
$e_{n+2}+(-1)^{n}b_{12}^{1}e_{1}-\sum\limits_{i=1}^{n+1}(-1)^{n+1-i}b_{1i}^{2}e_{i}$
for $e_{n+2}$ in (5), we get
 \vspace{2mm}\noindent
$\begin{array}{ll}
 (5)' ~\left\{\begin{array}{l}
{[}e_{1}, e_{3}, \cdots, e_{n+1}] = e_{1},\\
{[}e_{2}, \cdots, e_{n+1}] = e_{2},\\
{[}e_{2}, \cdots, \hat{e_{j}}, \cdots, e_{n+2}] = b_{1j}^{1}e_{1}, ~3\leq j\leq n+1,\\
{[}e_{1}, e_{3}, \cdots, \hat{e_{j}}, \cdots, e_{n+2}] =
b_{2j}^{1}e_{1}+b_{2j}^{2}e_{2}, ~3\leq j\leq n+1.
\end{array}\right.
\end{array} $

\vspace{2mm} We discuss $(5)'$ in two steps:

Step 1.  If $b_{1j}^{1}=0$ for $3\leq j\leq n+1$, then (5) is
isomorphic to

 \vspace{2mm}\noindent
 $\begin{array}{ll}  (5)''~
\left\{\begin{array}{l}
{[}e_{1}, e_{3}, \cdots, e_{n+1}] = e_{1},\\
{[}e_{2}, e_{3}, \cdots, e_{n+1}] = e_{2},\\
{[}e_{1}, e_{3}, \cdots, \hat{e_{j}}, \cdots, e_{n+1}, e_{n+2}] =
b_{2j}^{1}e_{1}+b_{2j}^{2}e_{2},(3\leq j\leq n+1).
\end{array}\right.
\end{array}$

For every $j$, $4\leq j\leq n+1$,  since

 \vspace{2mm} $ b_{23}^{1}[e_{1}, e_{3}, \cdots, \hat{e_{j}},\cdots, e_{n+2}]
 =[[e_{1}, e_{4}, \cdots, e_{n+2}], e_{3}, \cdots, \hat{e_{j}},\cdots, e_{n+2}]$

 \vspace{2mm}\noindent $=[b_{2j}^{1}e_{1}+b_{2j}^{2}e_{2}, e_{4}, \cdots, e_{n+2}]
 =b_{2j}^{1}[e_{1}, e_{4}, \cdots, e_{n+2}],$
\\we have $b_{23}^{1}b_{2j}^{2}=b_{23}^{2}b_{2j}^{1}.$ Then
$[e_{1}, e_{4}, \cdots, e_{n+2}]$ and $[e_{1}, e_{3}, \cdots,
\hat{e_{j}}, \cdots, e_{n+2}]$, $4\leq j \leq n+1$ are linearly
dependent.
\par
 If $[e_{1}, e_{3}, \cdots, \hat{e_{j}}, \cdots,
e_{n+2}]=0$ for $3\leq j\leq n+1$, then (5) is isomorphic to

\vspace{2mm}\noindent $\begin{array}{ll} (c^{7})~
\left\{\begin{array}{l}
{[}e_{1}, e_{3}, \cdots, e_{n+1}] = e_{1},\\
{[}e_{2}, e_{3}, \cdots, e_{n+1}] = e_{2}.
\end{array}\right.
\end{array}$

If there exists $[e_{1}, e_{3}, \cdots, \hat{e_{j}}, \cdots,
e_{n+2}]\neq 0$ for some $3\leq j\leq n+1$. We might as well suppose
$[e_{1}, e_{4}, \cdots, e_{n+2}]\neq 0$. Suppose $[e_{1}, e_{3},
\cdots, \hat{e_{j}}, \cdots, e_{n+2}]=k_{j}[e_{1}, e_{4}, \cdots,
e_{n+2}],$ for
 $4\leq j \leq n+1.$
 Substituting $e_{3}-\sum\limits_{j=4}^{n+1}(-1)^{i-4}k_{j}e_{j}$
for $e_{3}$ respectively in $(5)''$, we get

\vspace{2mm}\noindent $\begin{array}{ll} (5)'''~
\left\{\begin{array}{l}
{[}e_{1}, e_{3}, \cdots, e_{n+1}] = e_{1},\\
{[}e_{2}, e_{3}, \cdots, e_{n+1}] = e_{2},\\
{[}e_{1}, e_{4}, \cdots, e_{n+1}, e_{n+2}] =
b_{23}^{1}e_{1}+b_{23}^{2}e_{2},
\end{array}\right.
\end{array}$
where $b_{23}^{1}\neq 0$ or $b_{23}^{2}\neq 0.$ By suitable linear
transformations for the basis $e_1, \cdots, e_{n+2}$, we get
$(5)'''$ is isomorphic to $(c^4)$ or $(c^2)$ when $b_{23}^{1}\neq 0$
or $b_{23}^{2}\neq 0$ respectively.

Step 2. If there exists $b_{1j}^{1}\neq 0$ for some $3\leq j\leq
n+1$. The we might as well suppose $b_{13}^{1}\neq 0$. Substituting
$e_{3}-\sum\limits_{j=4}^{n+1}(-1)^{j-4}\frac{b_{1j}^{1}}{b_{13}^{1}}e_{j}$
and $\frac{1}{b_{13}^{1}}e_{n+2} $ for $e_{3}$ and $e_{n+2}$ in
$(5)'$ respectively, we get $(5)$ is isomorphic to

\vspace{2mm}\noindent
 $\begin{array}{ll}  ~
\left\{\begin{array}{l}
{[}e_{1}, e_{3}, \cdots, e_{n+1}] = e_{1},\\
{[}e_{2}, e_{3}, \cdots, e_{n+1}] = e_{2},\\
{[}e_{2}, e_{4}, \cdots, e_{n+1}, e_{n+2}] = e_{1},\\
{[}e_{1}, e_{3}, \cdots,\hat{e_{j}},\cdots, e_{n+1}, e_{n+2}] =
b_{2j}^{1}e_{1}+b_{2j}^{2}e_{2}, ~3\leq j\leq n+1.
\end{array}\right.
\end{array}$

The discussions is completely similar to the Step 1,  (5) is
isomorphic to $(c^2)$, $(c^4)$ or $(c^6)$ for the cases that
$b_{2j}^{1}, b_{2j}^{2}$ being to zero simultaneously or not.

Now we prove that $(c^i)$ is not isomorphic to $(c^j)$ when $i\neq
j$ for $1\leq i, j\leq 7.$ The case $(c^1)$ is not isomorphic to
$(c^3), (c^5)$ and $(c^7)$ since it is indecomposable. By Lemma 3.1
$(c^i)$ is not isomorphic to $(c^j)$ when $i\neq j$ for $i, j=3, 5,
7$. And $(c^j)$ for $j=1, 3, 5, 7$ are not isomorphic to $(c^2),
(c^4), (c^6)$ since they have nonzero center.

For the cases $(c^i), i=2, 4, 6$, we have Lie algebras $A_i=A$ (as
vector spaces) for $i=2, 4, 6$ respectively with products $[ , ]_1$
as follows

\vspace{2mm}\noindent $\begin{array}{ll} (c^{2})_1~
\left\{\begin{array}{l}
{[}e_{2}, e_3]_1 =  e_{1}, \\
{[}e_{2}, e'_{n+2}]_1 = e_{2},\\
{[}e_{1}, e'_{n+2}]_1 = e_{1};\\
\end{array}\right. ~(c^{4})_1~ \left\{\begin{array}{l}
{[}e_{2}, e_3]_1 = e_{1},\\
{[}e_{1}, e_{3}]_1 = e_{2}, \\
{[}e_{2}, e'_{n+2}]_1 = e_{2},\\
{[}e_{1}, e_{n+2}]_1 = e_{1};
\end{array}\right.
~ (c^{6})_1~ \left\{\begin{array}{l}
{[}e_{2}, e_3]_1=\alpha e_{1}+e_{2},\\
{[}e_{1}, e_{3}]_1= e_{2}, \\
{[}e_{2}, e'_{n+2}]_1= e_{2},\\
{[}e_{1}, e'_{n+2}]_1 = e_{1};
\end{array}\right.
\end{array}$

\vspace{2mm}\noindent  where $[x, y]_1=[x, y, e_4, \cdots, e_{n+1}]$
for $x, y\in A$ and $e'_{n+2}=(-1)^{n}e_{n+2}.$ And $A_i$ has
decomposition $A_i=Z(A_i)\dot+B_i $ (the direct sum as ideals),
where $B_i=Fe_1+Fe_2+Fe_3+Fe_{n+2}$ for $i=2, 4, 6$ are
$4$-dimensional solvable Lie algebras with multiplication table
$(c^i)_1$ respectively.

 It is easy to see that $H=Fe_3+\cdots+Fe_{n+2}$ is a Cartan subalgebra
of $(c^2), (c^4)$ and $(c^6)$ and the vectors $e_4, \cdots, e_{n+1}$
have the symmetric status in the multiplication. Then $(c^i)$ is
isomorphic to $(c^j)$ if and only if the Lie algebra $(c^i)_1$ is
isomorphic to $(c^j)_1$. By the classification [23] of
$4$-dimensional solvable Lie algebras, $(c^i)_1$ is not isomorphic
to $(c^j)_1$ for $i\neq j$. Then we get $(c^i)$ is not isomorphic to
$(c^j)$ when $i\neq j.$ And the $n$-Lie algebra of the case $(c^6)$
with coefficient $\alpha$ is isomorphic to that with coefficient
$\alpha'$ if and only if $\alpha=\alpha'.$

Summarizing, we get that $(c^i)$ is not isomorphic to $(c^j)$ if
$i\neq j$ for $1\leq i, j\leq 7.$

4. ~Let $\dim A^{1}=3$ and $A^{1}=Fe_{1}+Fe_{2}+Fe_{3}$. By Lemma
3.1, Lemma 3.2 and Lemma 3.3, the multiplication table of $A$ in a
basis $e_1, \cdots, e_{n+2}$ has only following possibilities:

\vspace{2mm}\noindent $\begin{array}{ll} (1)~
\left\{\begin{array}{l}
{[}e_{2}, \cdots, e_{n+1}] = e_{1},\\
{[}e_{1}, \cdots, \hat{e_{i}}, \cdots, \hat{e_{j}}, \cdots, e_{n+1},
e_{n+2}] = b_{ij}^{1}e_{1}+b_{ij}^{2}e_{2}+b_{ij}^{3}e_{3};
\end{array}\right.
\end{array}$

\vspace{2mm}\noindent $\begin{array}{ll} (2)~
\left\{\begin{array}{l}
{[}e_{1}, \cdots, e_{n}] = e_{1},\\
{[}e_{1}, \cdots, \hat{e_{i}}, \cdots, \hat{e_{j}}, \cdots, e_{n+1},
e_{n+2}] = b_{ij}^{1}e_{1}+b_{ij}^{2}e_{2}+b_{ij}^{3}e_{3};
\end{array}\right.
\end{array}$

\vspace{2mm}\noindent $\begin{array}{ll} (3)~
\left\{\begin{array}{l}
{[}e_{2}, \cdots, e_{n+1}] = e_{1},\\
{[}e_{1}, e_{3}, \cdots, e_{n+1}] = e_{2},\\
{[}e_{1}, \cdots, \hat{e_{i}}, \cdots, \hat{e_{j}}, \cdots, e_{n+1},
e_{n+2}] = b_{ij}^{1}e_{1}+b_{ij}^{2}e_{2}+b_{ij}^{3}e_{3};
\end{array}\right.
\end{array}$

\vspace{2mm}\noindent $\begin{array}{ll} (4)~
\left\{\begin{array}{l}
{[}e_{2}, \cdots, e_{n+1}] =\alpha e_{1}+ e_{2},\\
{[}e_{1}, e_{3}, \cdots, e_{n+1}] = e_{2},\\
{[}e_{1}, \cdots, \hat{e_{i}}, \cdots, \hat{e_{j}}, \cdots, e_{n+1},
e_{n+2}] = b_{ij}^{1}e_{1}+b_{ij}^{2}e_{2}+b_{ij}^{3}e_{3};
\end{array}\right.
\end{array}$

\vspace{2mm}\noindent $\begin{array}{ll} (5)~
\left\{\begin{array}{l}
{[}e_{1}, e_{3}, \cdots, e_{n+1}] =e_{1},\\
{[}e_{2}, e_{3}, \cdots, e_{n+1}] = e_{2},\\
{[}e_{1}, \cdots, \hat{e_{i}}, \cdots, \hat{e_{j}}, \cdots, e_{n+1},
e_{n+2}] = b_{ij}^{1}e_{1}+b_{ij}^{2}e_{2}+b_{ij}^{3}e_{3};
\end{array}\right.
\end{array}$

\vspace{2mm}\noindent $\begin{array}{ll} (6)~
\left\{\begin{array}{l}
{[}e_{2}, \cdots, e_{n+1}] = e_{1},\\
{[}e_{1}, e_{3}, \cdots, e_{n+1}] = e_{2},\\
{[}e_{1}, e_{2}, e_{4}, \cdots, e_{n+1}] = e_{3},\\
{[}e_{1}, \cdots, \hat{e_{i}}, \cdots, \hat{e_{j}}, \cdots, e_{n+1},
e_{n+2}] = b_{ij}^{1}e_{1}+b_{ij}^{2}e_{2}+b_{ij}^{3}e_{3};
\end{array}\right.
\end{array}$

\vspace{2mm}\noindent $\begin{array}{ll} (7)~\left\{\begin{array}{l}
{[}e_{3}, e_{4}, \cdots, e_{n+2}]=b_{12}^{1}e_{1}+b_{12}^{2} e_{2}+b_{12}^{3}e_{3},\\
{[}e_{2}, e_{4}, \cdots, e_{n+2}]=b_{13}^{1}e_{1}+b_{13}^{2} e_{2}+b_{13}^{3}e_{3},\\
{[}e_{1}, e_{4}, \cdots,
e_{n+2}]=b_{23}^{1}e_{1}+b_{23}^{2}e_{2}+b_{23}^{3}e_{3},
\end{array}\right.
\end{array}$

\vspace{2mm}\noindent where $b_{ij}\in F,1\leq i< j\leq n+1$.

Firstly, we study the case (1). Substituting the first identity into
the other equations, we get

\vspace{2mm}\noindent $\sum\limits_{k=1}^{3}b_{ij}^{k}e_{k}=[e_{1},
e_{2}, e_{3}, \cdots, \hat{e_{i}}, \cdots, \hat{e_{j}}, \cdots,
e_{n+1}, e_{n+2}]$

\vspace{2mm}\noindent $=[[e_{2}, \cdots, e_{n+1}], e_{2}, e_{3},
\cdots, \hat{e_{i}}, \cdots, \hat{e_{j}}, \cdots, e_{n+1}, e_{n+2}]$

\vspace{2mm}\noindent $=[e_{2}, e_{3}, \cdots,
(-1)^{i-2}\sum\limits_{k=1}^{3}b_{1j}^{k}e_{k}, \cdots, e_{n+1}]
+[e_{2}, e_{3}, \cdots,
(-1)^{j-3}\sum\limits_{k=1}^{3}b_{1i}^{k}e_{k}, \cdots, e_{n+1}]$

\vspace{2mm}\noindent $=0,$ for $4\leq i < j\leq n+1,$

\vspace{2mm}\noindent $\sum\limits_{k=1}^{3}b_{2j}^{k}e_{k}=[e_{1},
e_{3}, \cdots, \hat{e_{j}}, \cdots, e_{n+1}, e_{n+2}]$

\vspace{2mm}\noindent $=[[e_{2}, \cdots, e_{n+1}], e_{3}, \cdots,
\hat{e_{j}}, \cdots, e_{n+1}, e_{n+2}]=b_{1j}^{2}e_{1},$ for $4\leq
j\leq n+1,$

\vspace{2mm}\noindent $\sum\limits_{k=1}^{3}b_{3j}^{k}e_{k}=[e_{1},
e_{2}, e_4, \cdots, \hat{e_{j}}, \cdots, e_{n+1}, e_{n+2}]$

\vspace{2mm}\noindent $=[[e_{2}, \cdots, e_{n+1}], e_{2}, \cdots,
\hat{e_{j}}, \cdots, e_{n+1}, e_{n+2}]=-b_{1j}^{3}e_{1},$ for $4\leq
j\leq n+1,$

\vspace{2mm}\noindent $\sum\limits_{k=1}^{3}b_{23}^{k}e_{k}=[e_{1},
e_{4}, \cdots, e_{n+2}]$

\vspace{2mm}\noindent $=[[e_{2}, \cdots, e_{n+1}], e_{4}, \cdots,
e_{n+2}]=b_{13}^{2}e_{1}+b_{12}^{3}e_{1}.$

\vspace{2mm}If we replace
$e_{n+2}-\sum\limits_{j=2}^{n+1}(-1)^{n+1-j}b_{1j}^{1}e_{j}$ for
$e_{n+2}$, then (1) is isomorphic to

\vspace{2mm}\noindent $\begin{array}{ll} (1)'
~\left\{\begin{array}{l}
{[}e_{2}, \cdots, e_{n+1}] = e_{1},\\
{[}e_{3}, \cdots, e_{n+2}] = \sum\limits_{k=2}^{3}b_{12}^{k}e_{k},\\
{[}e_{2}, e_{4}, \cdots, e_{n+2}] = \sum\limits_{k=2}^{3}b_{13}^{k}e_{k},\\
{[}e_{2}, e_{3}, \cdots, \hat{e_{j}}, \cdots, e_{n+1}, e_{n+2}]
= \sum\limits_{k=2}^{3}b_{1j}^{k}e_{k},~4\leq j \leq n+1,\\
{[}e_{1}, e_{4}, e_{5}, \cdots, e_{n+2}]=b_{23}^{1}e_{1}=(b_{13}^{2}+b_{12}^{3})e_{1},\\
{[}e_{1}, e_{3}, \cdots, \hat{e_{j}}, \cdots, e_{n+1}, e_{n+2}]
 =b_{2j}^{1}e_{1}=b_{1j}^{2}e_{1}, ~4\leq j \leq n+1,\\
{[}e_{1}, e_{2}, \cdots, \hat{e_{j}}, \cdots, e_{n+1}, e_{n+2}]
=b_{3j}^{1}e_{1}=-b_{1j}^{3}e_{1},~4\leq j \leq n+1.
\end{array}\right.
\end{array}$

\vspace{2mm} Fixing $e_{n+2}$ in the $n$-ary multiplication of $A$,
we get an $(n+2)$-dimensional $(n-1)$-Lie algebra $A_0=A$ (as vector
space) with production $[ , \cdots, ]_0$ and the multiplication
table of $A_0$ in the basis $e_1, \cdots, e_{n+2}$ is as follows

\vspace{2mm}\noindent $\begin{array}{ll}
 \left\{\begin{array}{l}
{[}e_{3}, e_4, \cdots, e_{n+1}]_0 = \sum\limits_{k=2}^{3}b_{12}^{k}e_{k},\\
{[}e_{2}, e_{4}, \cdots, e_{n+1}]_0 = \sum\limits_{k=2}^{3}b_{13}^{k}e_{k},\\
{[}e_{2}, e_{3}, \cdots, \hat{e_{j}}, \cdots, e_{n+1}]_0
= \sum\limits_{k=2}^{3}b_{1j}^{k}e_{k},~4\leq j \leq n+1,\\

{[}e_{1}, e_{4}, e_{5}, \cdots, e_{n+1}]_0=b_{23}^{1}e_{1}=(b_{13}^{2}+b_{12}^{3})e_{1},\\
{[}e_{1}, e_{3}, \cdots, \hat{e_{j}}, \cdots, e_{n+1}]_0
 =b_{2j}^{1}e_{1}=b_{1j}^{2}e_{1}, ~4\leq j \leq n+1,\\
{[}e_{1}, e_{2}, \cdots, \hat{e_{j}}, \cdots, e_{n+1}]_0
=b_{3j}^{1}e_{1}=-b_{1j}^{3}e_{1},~4\leq j \leq n+1.
\end{array}\right.
\end{array}$

\vspace{2mm}\noindent Set $B=Fe_2+\cdots+Fe_{n+1}$. Then $B$ is a
subalgebra of $A_0$, $\dim B^1=2$ since $\dim A_0^1=\dim A^1=3,$ and
the multiplication table of $B$ in the basis $e_2, \cdots, e_{n+1}$
is as follows

\vspace{2mm}\noindent $\begin{array}{ll}
 ~\left\{\begin{array}{l}
{[}e_{3}, e_{4}, \cdots, e_{n+1}]_{0} = b_{12}^{2}e_{2}+b_{12}^{3}e_{3},\\
{[}e_{2}, e_{4}, \cdots, e_{n+1}]_{0} = b_{13}^{2}e_{2}+b_{13}^{3}e_{3},\\
{[}e_{2}, e_{3}, e_{4}, \cdots, \hat{e_{j}}, \cdots, e_{n+1}]_{0} =
b_{1j}^{2}e_{2}+b_{1j}^{3}e_{3}, ~4\leq j \leq n+1.
\end{array}\right.
\end{array}$

\vspace{2mm}\noindent By discussions completely similar  to [3], we
have

\vspace{2mm}\noindent $\Delta=\left|
\begin{array}{ccc}
b_{12}^2 & b_{12}^3 \\
 b_{13}^2 & b_{13}^3
\end{array}
\right|\neq0, ~ \mbox{and}~ b_{1j}^{2}=b_{1j}^{3}=0, ~\mbox{for}
~4\leq j \leq n.$

\vspace{2mm}\noindent Therefore $(1)'$ has the form

\vspace{2mm}\noindent $\begin{array}{ll} (1)''
~\left\{\begin{array}{l}
{[}e_{2}, \cdots, e_{n+1}] = e_{1},\\
{[}e_{3}, \cdots, e_{n+2}] = b_{12}^{2}e_{2}+b_{12}^{3}e_{3},\\
{[}e_{2}, e_{4}, \cdots, e_{n+2}] = b_{13}^{2}e_{2}+b_{13}^{3}e_{3},\\
{[}e_{1}, e_{4}, e_{5}, \cdots,
e_{n+2}]=(b_{13}^{2}+b_{12}^{3})e_{1},
\end{array}\right.
\end{array} ~\mbox{where}~ \Delta=\left|
\begin{array}{ccc}
b_{12}^2 & b_{12}^3 \\
 b_{13}^2 & b_{13}^3
\end{array}
\right|\neq0.$

\vspace{2mm}If $b_{13}^{2}+b_{12}^{3}=0$, and $b_{12}^{2}\neq0$,
taking a linear transformation for basis $e_1, \cdots, e_{n+2}$ by
replacing $\frac{2\sqrt{\Delta}}{b_{12}^2}e_1$ for $e_1$,
$e_{2}+\frac{b_{12}^{3}-\sqrt{\Delta}}{ b_{12}^{2}}e_{3}$ for
$e_{2}$, $e_{2}+\frac{b_{12}^{3}+\sqrt{\Delta}}{ b_{12}^{2}}e_{3}$
for $e_3$ and $\frac{1}{\sqrt{\Delta}}e_{n+2}$ for $e_{n+2}$ in
$(1)''$, we get that $(1)$ is isomorphic to

\vspace{2mm}\noindent $\begin{array}{ll} (d^{1})~
\left\{\begin{array}{l}
{[}e_{2}, \cdots, e_{n+1}] = e_{1},\\
{[}e_{2}, e_{4}, \cdots, e_{n+2}]=-e_{2},\\
{[}e_{3}, \cdots, e_{n+2}] =e_{3}.
\end{array}\right.
\end{array}$

\vspace{2mm}In the case that $b_{13}^{2}+b_{12}^{3}=b_{12}^{2}=0$,
by discussions similar
 to above we get $(1)$ is isomorphic to $(d^1).$

If $b_{13}^{2}+b_{12}^{3}\neq 0$,
 and $b_{12}^{2}\neq0$, taking a linear transformation for basis $e_1, \cdots, e_{n+2}$
 by replacing
$\frac{\Delta}{b_{12}^2(b_{13}^2+b_{12}^3)}e_1$ for $e_1$,
$e_2+\frac{b_{12}^3}{b_{12}^2}e_3$ for $e_2$,
$e_2+\frac{1}{b_{12}^2}(b_{12}^3+\frac{\Delta}{b_{13}^2+b_{12}^3})e_3$
for $e_3$, $\frac{1}{b_{13}^2+b_{12}^3}e_{n+2}$ for $e_{n+2}$ in
$(1)''$,  we get $(1)$ isomorphic to

\vspace{2mm}\noindent $\begin{array}{ll} (d^{2}) ~
\left\{\begin{array}{l}
{[}e_{2}, \cdots, e_{n+1}]=e_{1}, \\
{[}e_{3}, \cdots, e_{n+2}]=e_{3}+\alpha e_{2}, \\
{[}e_{2}, e_{4}, \cdots, e_{n+2}]=e_{3}, \\
{[}e_{1}, e_{4}, \cdots, e_{n+2}]=e_{1},
\end{array}\right.
\end{array}
~\mbox{where} ~\alpha=\frac{\Delta}{(b_{13}^2+b_{12}^3)^{2}}\in F
~\mbox{and}~\alpha\neq0.$

\vspace{2mm}If $b_{13}^{2}+b_{12}^{3}\neq 0$
 and $b_{12}^{2}=0$, $(1)''$ is of the form

\vspace{2mm}\noindent $\begin{array}{ll} (1)'''
~\left\{\begin{array}{l}
{[}e_{2}, \cdots, e_{n+1}] = e_{1},\\
{[}e_{3}, \cdots, e_{n+2}] = b_{12}^{3}e_{3},\\
{[}e_{2}, e_{4}, \cdots, e_{n+2}] = b_{13}^{2}e_{2}+b_{13}^{3}e_{3},\\
{[}e_{1}, e_{4}, e_{5}, \cdots,
e_{n+2}]=(b_{13}^{2}+b_{12}^{3})e_{1}.
\end{array}\right.
\end{array}$

\vspace{2mm}\noindent By similar discussions as above, taking
suitable linear transformation for basis $e_1,$ $\cdots,$ $e_{n+2}$,
we get $(1)'''$ is isomorphic to $(d^2)$ in the case of
$b_{13}^{3}\neq 0$ or  in the case of $b_{13}^{3}=0$ and
$b_{13}^{2}\neq b_{12}^{3}$. In the case of $b_{13}^{3}=0$ and
$b_{13}^{2}=b_{12}^{3}$, it is evident that $(1)'''$ is of the form

\vspace{2mm}\noindent $\begin{array}{ll}
 (d^{3})~ \left\{\begin{array}{l}
{[}e_{2}, \cdots, e_{n+1}]=e_{1}, \\
{[}e_{3}, e_{4}, \cdots, e_{n+2}]=e_{3}, \\
{[}e_{2}, e_{4}, \cdots, e_{n+2}]=e_{2}, \\
{[}e_{1}, e_{4}, \cdots, e_{n+2}]=2e_{1}.
\end{array}\right.
\end{array}$

Secondly substituting
$e_{n+2}-\sum\limits_{i=1}^{n}(-1)^{n-i}b_{in+1}^{1}e_{i}$ for
$e_{n+2}$ in (2), we get

\vspace{2mm}\noindent $\begin{array}{ll} (2)'
~\left\{\begin{array}{l}
{[}e_{1}, \cdots, e_{n}] = e_{1},\\
{[}e_{1}, \cdots, \hat{e_{i}}, \cdots, \hat{e_{j}}, \cdots, e_{n},
e_{n+1}, e_{n+2}]
=\sum\limits_{k=1}^{3}b_{ij}^{k}e_{k}, ~ 1\leq i<j \leq n,\\
{[}e_{1}, \cdots, \hat{e_{i}}, \cdots, e_{n},
e_{n+2}]=\sum\limits_{k=2}^{3}b_{in+1}^{k}e_{k},
~ 1\leq i \leq n.\\
\end{array}\right.
\end{array}$

Using the Jacobi identities for $\{[e_{1},\cdots,~ e_{n}],e_{3},$ $
\cdots, e_{n}, e_{n+2}\},$ ~$\{[e_{1},\cdots,~ e_{n}],$ $ e_{2},
e_{4}, \cdots,$ $ e_{n}, e_{n+2}\}$, ~$\{[e_{1},\cdots,~ e_{n}],$ $
e_{2}, e_{3}, e_{4},$ $ \cdots, \hat{e_{i}},\cdots, e_{n},
e_{n+2}\}$ for $4\leq i\leq n$, ~$\{[e_{1},e_{2},e_{4},$ $\cdots,
e_{n},$ $e_{n+2}],$ $e_{2},\cdots,$ $~\hat{e_{j}},\cdots,~
e_{n+1}\}$ for $2\leq j\leq n$, ~$\{ [e_{2},~  \cdots,~ e_{n+1}] ,$
$e_{2},\cdots,$ $\hat{e_{i}},$ $\cdots,$ $~\hat{e_{j}},$
$\cdots,e_{n}, e_{n+1},e_{n+2}]$ for $2\leq i\leq j\leq n$, we get
$b_{ij}^{3}=0$ for $1\leq i<j \leq n$. Hence (2) is of the form

\vspace{2mm}\noindent $\begin{array}{ll} \left\{\begin{array}{l}
{[}e_{1}, \cdots, e_{n}] = e_{1},\\
{[}e_{1}, \cdots, \hat{e_{i}}, \cdots, \hat{e_{j}}, \cdots, e_{n},
e_{n+1}, e_{n+2}]=\sum\limits_{k=1}^{2}b_{ij}^{k}e_{k}, ~1\leq i<j
\leq n.
\end{array}\right.
\end{array}$

\vspace{2mm}\noindent This contradicts $\dim A^1=3.$ Therefore the
case (2) is not realized.

Thirdly, imposing the Jacobi identities on (3) for $\{[e_{2},
\cdots, e_{n+1}],$ $ e_{2}, \cdots, \hat{e_{i}},
\cdots,~\hat{e_{j}},$ $\cdots, e_{n},$ $ e_{n+1}, e_{n+2}\}$ for
$2\leq i<j\leq n+1$, $\{[e_{1},e_{3},~  \cdots,~ e_{n+1}],$ $e_{3},
 \cdots, $ $~\hat{e_{j}}, \cdots,$ $
e_{n+1},e_{n+2}\}$ for $ 3\leq j\leq n+1$, $\{[e_{1},e_{3},~
\cdots,~ e_{n+1}],e_{4},\cdots, e_{n+2}\}$ and $\{[e_{2},~ \cdots,~
e_{n+1}],$ $e_{4},$ $\cdots, e_{n+2}\}$,
 we get $b_{ij}^{3}=0$ for $2\leq i<j\leq n+1$, $b_{1j}^{3}=0$ for $ 3\leq j\leq n+1$,
$b_{12}^{3}=0$ respectively. Then (3) is of the form

\vspace{2mm}\noindent $\begin{array}{ll} \left\{\begin{array}{l}
{[}e_{2}, \cdots, e_{n+1}] = e_{1},\\
{[}e_{1}, e_{3}, \cdots, e_{n}] = e_{2},\\
{[}e_{1}, \cdots, \hat{e_{i}}, \cdots, \hat{e_{j}}, \cdots, e_{n+1},
e_{n+2}]=\sum\limits_{k=1}^{2}b_{ij}^{k}e_{k}, ~1\leq i<j \leq n+1.
\end{array}\right.
\end{array}$

\vspace{2mm}\noindent This contradicts $\dim A^1=3.$ Therefore, the
case (3) is not realized.

The cases (4) and (5) are not realized by discussions similar to the
case (3).

Fourthly, for $4\leq i<j\leq n+1$, from the table (6)

\vspace{2mm}\noindent $\sum\limits_{k=1}^{3}b_{ij}^{k}e_{k}=[e_{1},
e_{2}, e_{3}, e_{4}, \cdots, \hat{e_{i}}, \cdots, \hat{e_{j}},
\cdots, e_{n+1}, e_{n+2}]$

\vspace{2mm}\noindent $=[[e_{2}, \cdots, e_{n+1}], e_{2}, e_{3},
e_{4}, e_{5}, \cdots,
 \hat{e_{i}}, \cdots, \hat{e_{j}}, \cdots, e_{n+1}, e_{n+2}]$

\vspace{2mm}\noindent $=[e_{2}, e_{3}, e_{4}, \cdots,
(-1)^{i-2}\sum\limits_{k=1}^{3}b_{1j}^{k}e_{k}, \cdots, e_{n+1}]
$

\vspace{2mm}\noindent $+[e_{2}, e_{3}, e_{4}, \cdots,
(-1)^{j-3}\sum\limits_{k=1}^{3}b_{1i}^{k}e_{k}, \cdots, e_{n+1}]=0.$

\vspace{2mm}\noindent Then (6) has the form

\vspace{2mm}\noindent $\begin{array}{ll} (6)'
~\left\{\begin{array}{l}
{[}e_{2}, \cdots, e_{n+1}] = e_{1},\\
{[}e_{1}, e_{3}, \cdots, e_{n}] = e_{2},\\
{[}e_{1}, e_{2}, e_{4}, \cdots, e_{n+1}] = e_{3},\\
{[}e_{1}, \cdots, \hat{e_{i}}, \cdots, \hat{e_{j}}, \cdots, e_{n+1},
e_{n+2}]=\sum\limits_{k=1}^{3}b_{ij}^{k}e_{k}, ~1\leq i \leq 3, i<j
\leq n+1.
\end{array}\right.
\end{array}$

\vspace{2mm}\noindent For $4\leq j\leq n+1$, imposing the Jacobi
identities for $\{[e_{1}, e_{3}, \cdots, e_{n+1}],$ $e_{3},
e_{4},\cdots,$ $~\hat{e_{j}}, \cdots,$ $~ e_{n+1},$ $~ e_{n+2}\}$,
$\{e_{2},$ $[e_{1},e_{2},e_{4},$ $\cdots,~ e_{n+1}],$
$e_{4},\cdots,$ $~\hat{e_{j}},\cdots,$ $e_{n+1},~ e_{n+2}\},$
$\{[e_{2}, \cdots, e_{n+1}],$ $e_{3}, e_{4},$ $\cdots,
~\hat{e_{j}},$ $\cdots,~ e_{n+1},~ e_{n+2}\}$, $\{e_{1}, [e_{1},
e_{2}, e_{4},$ $\cdots,~ e_{n+1}],$ $e_{4},$ $\cdots,$
$\hat{e_{j}},$ $\cdots,~ e_{n+1},~ e_{n+2}\}$, $\{[e_{2},\cdots,~
e_{n+1}],$ $e_{2},$ $e_{4},\cdots,$ $~\hat{e_{j}},$ $\cdots,~
e_{n+1},~ e_{n+2}\},$ $\{e_{1},$ $[e_{1},e_{3}, \cdots,$ $~
e_{n+1}],$ $e_{4}, \cdots,$ $~\hat{e_{j}},$ $\cdots,~ e_{n+1},$ $~
e_{n+2}\},$ $\{[e_{1},e_{2},$ $e_{4},\cdots, e_{n+1}],$ $e_{4},$
$\cdots, e_{n+2}\}$, $\{[e_{1}, e_{3}, \cdots, e_{n+1}],$ $e_{4},$ $
\cdots,~ e_{n+2}\},$ $\{[e_{2},\cdots, e_{n+1}],$ $e_{4},$ $\cdots,~
e_{n+2}\}$,
 we get
$b_{1j}^{3}=0,$ $b_{1j}^{1}=b_{2j}^{2},$ $b_{1j}^{2}=b_{2j}^{1};$
$b_{1j}^{2}=0,$ $b_{1j}^{3}=-b_{3j}^{1},$ $b_{1j}^{1}=b_{3j}^{3};$
$b_{2j}^{3}=0,$ $b_{2j}^{1}=b_{1j}^{2},$ $b_{2j}^{2}=b_{1j}^{1}$;
$b_{2j}^{1}=0,$ $b_{2j}^{2}=b_{3j}^{3},$ $b_{2j}^{3}=b_{3j}^{2}$;
$b_{3j}^{2}=0,$ $b_{3j}^{1}=-b_{1j}^{3},$ $b_{3j}^{3}=b_{1j}^{1};$ $
b_{3j}^{1}=0,$ $b_{3j}^{2}=b_{2j}^{3},$ $b_{3j}^{3}=b_{2j}^{2}$;
$b_{12}^{1}=-b_{23}^{3}, $ $b_{12}^{2}=b_{13}^{3},$
$b_{12}^{3}=b_{13}^{2}+b_{23}^{1}$; $b_{13}^{1}=b_{23}^{2},$
$b_{13}^{2}=b_{12}^{3}+b_{23}^{1},$ $b_{13}^{3}=b_{12}^{2}$; $
b_{23}^{1}=b_{12}^{3}+b_{13}^{2},$ $b_{23}^{2}=b_{13}^{1},$
$b_{23}^{3}=-b_{12}^{1}$ respectively. Then $(6)'$ is of the form

\vspace{2mm}\noindent $\begin{array}{ll} \left\{\begin{array}{l}
{[}e_{2}, \cdots, e_{n+1}] = e_{1},\\
{[}e_{1}, e_{3}, \cdots, e_{n+1}] = e_{2},\\
{[}e_{1}, e_{2}, e_{4}, \cdots, e_{n+1}] = e_{3},\\
{[}e_{3}, \cdots, e_{n+2}] = b_{12}^{1}e_{1}+ b_{12}^{2}e_{2},\\
{[}e_{2}, e_{4}, \cdots, e_{n+2}] = b_{13}^{1}e_{1}+ b_{12}^{2}e_{3},\\
{[}e_{1}, e_{4}, e_{5}, \cdots, e_{n+2}]= b_{13}^{1}e_{2}-b_{12}^{1}e_{3},\\
{[}e_{2}, e_{3}, e_{4}, \cdots, \hat{e_{j}}, \cdots, e_{n+1}, e_{n+2}] = b_{1j}^{1}e_{1}, ~4\leq j \leq n+1,\\
{[}e_{1}, e_{3}, e_{4}, \cdots, \hat{e_{j}}, \cdots, e_{n+1},~
e_{n+2}] =b_{2j}^{2}e_{2}=b_{1j}^{1}e_{2},
~4\leq j \leq n+1,\\
{[}e_{1}, e_{2}, e_{4}, \cdots, \hat{e_{j}}, \cdots, e_{n+1},
e_{n+2}] =b_{3j}^{3}e_{3}=b_{1j}^{1}e_{3}, ~4\leq j \leq n+1.
\end{array}\right.
\end{array}$

\vspace{2mm}\noindent Substituting
$e_{n+2}-\sum\limits_{j=2}^{n+1}(-1)^{n+1-j}b_{1j}^{1}e_{j}-(-1)^{n-1}b_{12}^{2}e_{1}$
for $e_{n+2}$, we get (6) is isomorphic to

\vspace{2mm}\noindent $\begin{array}{ll}
(d^{4})~\left\{\begin{array}{l}
{[}e_{2}, \cdots, e_{n+1}] = e_{1},\\
{[}e_{1}, e_{3}, \cdots, e_{n+1}] = e_{2},\\
{[}e_{1}, e_{2}, e_{4}, \cdots, e_{n+1}] = e_{3}.
\end{array}\right.
\end{array}$

\vspace{2mm} Lastly we discuss the case (7). It follows by a simple
computation that there does not exist any nonabelian proper
subalgebra of $A$ containing $A^1.$ Then the
 the multiplication of $A$ is completely determined
by the left multiplication ad$(e_{4}, \cdots, e_{n+2})$. And
ad$(e_{4}, \cdots, e_{n+2})|_{A^1}$ is  nonsingular since $\dim
A^1=3$. So we can choose a  basis $e_1, e_2, e_3$ of $A^1$ such that
the multiplication of $A$ in the basis $e_1, \cdots, e_{n+2}$ has
the following  possibilities

\vspace{2mm}\noindent $\begin{array}{l} (d^{5})' ~
\left\{\begin{array}{l}
{[}e_{1}, e_{4}, \cdots, e_{n+2}]=\beta_{1}e_{1}, \\
{[}e_{2}, e_{4}, \cdots, e_{n+2}]=\beta_{2}e_{2}, \\
{[}e_{3}, e_4, \cdots, e_{n+2}]=\beta_{3}e_{3},
\end{array}\right.
\end{array} ~\beta_{i}\in F, \beta_i\neq0, ~i=1, 2, 3;
$

\vspace{2mm}\noindent $
\begin{array}{l}
(d^{6})' ~ \left\{\begin{array}{l}
{[}e_{1}, e_{4}, \cdots, e_{n+2}]=\alpha e_{1}+e_{2}, \\
{[}e_{2}, e_{4}, \cdots, e_{n+2}]=\alpha e_{2}+e_{3}, \\
{[}e_{3}, e_4, \cdots, e_{n+2}]=\alpha e_{3}, \\
\end{array}\right. \alpha\in F, \alpha\neq 0;
\end{array}
$

\vspace{2mm}\noindent $\begin{array}{l}
 (d^{7})'  ~ \left\{\begin{array}{l}
{[}e_{1}, e_{4}, \cdots, e_{n+2}]=\gamma _{1} e_{1}+e_{2}, \\
{[}e_{2}, e_{4}, \cdots, e_{n+2}]=\gamma_{1} e_{2}, \\
{[}e_{3}, e_4, \cdots, e_{n+2}]=\gamma_{2} e_{3},
\end{array}\right.
\end{array} ~\gamma_j\in F, \gamma_j\neq 0, ~j=1, 2.$

\vspace{1mm}If we fix $e_5, \cdots, e_{n+2}$ in the $n$-ary
multiplication of $A$, we get solvable Lie algebra $A_1=A$ ( as
vector spaces) with the Lie production  $[ , ]_1$

 \vspace{2mm}\noindent $
 [x, y]_1=[x, y, e_5, \cdots, e_{n+2}], ~ x, y\in A_1.
 $

 \vspace{2mm}\noindent
Then the multiplication tables of $A_1$  with respect to $(d^5)',
(d^6),' (d^{7})'$  are

\vspace{2mm}\noindent  $\begin{array}{l} (d^{5})'' ~
\left\{\begin{array}{l}
{[}e_{1}, e_{4}]_1=\beta_{1}e_{1}, \\
{[}e_{2}, e_{4}]_1=\beta_{2}e_{2}, \\
{[}e_{3}, e_{4}]_1=\beta_{3}e_{3},
\end{array}\right.
\end{array} ~\beta_{i}\in F, \beta_i\neq0, 1=1, 2, 3;
$

\vspace{2mm}\noindent $
\begin{array}{l}
(d^{6})'' ~ \left\{\begin{array}{l}
{[}e_{1}, e_{4}]_1=\alpha e_{1}+e_{2}, \\
{[}e_{2}, e_{4}]_1=\alpha e_{2}+e_{3}, \\
{[}e_{3}, e_{4}]_1=\alpha  e_{3}, \\
\end{array}\right.
\end{array} \alpha\in F, \alpha\neq 0;
$

\vspace{2mm}\noindent $
\begin{array}{l}
(d^{7})''  ~ \left\{\begin{array}{l}
{[}e_{1}, e_{4}]_1=\gamma _{1}e_{1}+e_{2}, \\
{[}e_{2}, e_{4}]_1=\gamma_{1}e_{2}, \\
{[}e_{3}, e_{4}]_1=\gamma_{2}e_{3},
\end{array}\right.
\end{array} ~\gamma_j\in F, \gamma_j\neq 0, j=1, 2.
$

\vspace{2mm}\noindent This implies that $(d^i)''$ can be decomposed
into the direct sum of ideals $Z(A_1)$ and $B$, where the center
$Z(A_1)=Fe_5+\cdots+Fe_{n+2}$ and the ideal $B=Fe_1+Fe_2+Fe_3+Fe_4$,
$i=5, 6, 7$. By the classification of $4$-dimensional solvable Lie
algebras [23], we get that one and only one of following
possibilities holds up to isomorphisms:

\vspace{2mm}\noindent  $\begin{array}{l}
 (d^{5})~ \left\{\begin{array}{l}
{[}e_{1}, e_{4}, \cdots, e_{n+2}]=e_{1},\\
{[}e_{2}, e_{4}, \cdots, e_{n+2}]= e_{3},\\
{[}e_{3}, e_{4}, \cdots, e_{n+2}]= \beta e_2+(1+\beta)e_{3}, ~
\beta\in F, \beta\neq 0, 1;
\end{array}\right.
\end{array}$

\vspace{2mm}\noindent $ \begin{array}{l} (d^{6})~
\left\{\begin{array}{l}
{[}e_{1}, e_{4}, \cdots, e_{n+2}]=e_{1},\\
{[}e_{2}, e_{4}, \cdots, e_{n+2}]=e_{2},\\
{[}e_{3}, e_{4}, \cdots, e_{n+2}]=e_{3};
\end{array}\right.
\end{array}
$

\vspace{2mm}\noindent  $\begin{array}{l}
 (d^{7})~ \left\{\begin{array}{l}
{[}e_{1}, e_{4}, \cdots, e_{n+2}]=e_{2},\\
{[}e_{2}, e_{4}, \cdots, e_{n+2}]= e_{3},\\
{[}e_{3}, e_{4}, \cdots, e_{n+2}]= se_1+te_2+ue_{3}, ~s, t, u\in
F,~s\neq 0.
\end{array}\right.
\end{array}
$

\vspace{2mm} \noindent And $(d^i)$ is not isomorphic to $(d^j)$ when
$i\neq j$ for $ 5\leq i, ~j\leq 7.$ And the $n$-Lie algebras
corresponding to the case $(d^5)$ with coefficients $\beta$ and
$\beta'$ are isomorphic if and only if $\beta=\beta'.$  We also have
that the $n$-Lie algebras corresponding to the case $(d^{7})$ with
coefficients  $s, t, u$ and $s', t', u'$ are isomorphic  if and only
if there exists a nonzero element $r\in F$ such that
$$
s=r^3 s', ~t=r^2 t', ~u=ru', ~s, s', t, t', u, u' \in F.
$$

It is evident that $(d^{5}), (d^{6}), (d^{7})$ are not isomorphic
 to other cases since
$(d^{5}),$ $ (d^{6}), $ $(d^{7})$ have  no nonabelian proper
subalgebras containing $A^1$. The case $(d^{4})$ is not isomorphic
to any cases of $(d^{1}), (d^{2}), (d^{3})$ since $(d^{4})$ is
decomposable. Because $(d^{1})$ has non-trivial center, $(d^{1})$ is
not isomorphic to $(d^{2})$ and $(d^{3})$. By direct computation we
know that  dimensions of the derivation algebras of the case $(d^2)$
and $(d^3)$ are $n^2+2$ and $n^2+3$ respectively, therefore, $(d^2)$
and $d^3$ represent non-isomorphic classes.

Now fixing $e_4, \cdots, e_{n+1}$ in the multiplication of $A$ of
the case $(d^2)$, and substituting $(-1)^{n-2}e_{n+2}$ for
$e_{n+2}$, we get a solvable Lie algebra $A_2$ ($A_2=A$ as vector
spaces) with the product
$$[x , y]_2=[x, y, e_4, \cdots, e_{n+1}],
$$
and the multiplication table of $A_2$ in the basis $e_1, \cdots,
e_{n+2}$ is as follows
$$
\begin{array}{l}
\left\{\begin{array}{l}
{[}e_2, e_3]_2=e_1,\\
{[}e_3, e_{n+2}]_2=e_3+\alpha e_2,\\
{[}e_2, e_{n+2}]_2=e_2,\\
{[}e_1, e_{n+2}]_2=e_1.\\
\end{array}\right.
\end{array}
$$
Then $A_2$ has a decomposition  $A_2=B\oplus Z(A_2)$, where the
center $Z(A_2)=Fe_1+\cdots+Fe_{n+1},$  and
$B=Fe_1+Fe_2+Fe_3+Fe_{n+2}$ is an ideal. By the classification of
solvable Lie algebras [23], we get that $n$-Lie algebras of the case
$(d^2)$ with coefficients $\alpha$ and $\alpha'$ are isomorphic if
and only if $\alpha=\alpha'$.

5. By Lemma 3.3, we have $\dim A^{1}=r, ~4\leq r\leq n+1.$ Suppose
 $A^{1}=Fe_{1}+\ldots+Fe_{r}.$ From Lemma 3.1 and Lemma 3.2, the
 multiplication table of $A$ in the basis $e_1, \cdots, e_{n+2}$ has
 following possibilities

\vspace{2mm}\noindent $\begin{array}{ll} (1)~
\left\{\begin{array}{l}
{[}e_{2}, \cdots, e_{n+1}] = e_{1},\\
{[}e_{1}, \cdots, \hat{e_{i}}, \cdots, \hat{e_{j}}, \cdots, e_{n+1},
e_{n+2}] = \sum\limits_{k=1}^{r}b_{ij}^{k}e_{k};
\end{array}\right.
\end{array}$

\vspace{2mm}\noindent $\begin{array}{ll} (2)~
\left\{\begin{array}{l}
{[}e_{1}, \cdots, e_{n}] = e_{1},\\
{[}e_{1}, \cdots, \hat{e_{i}}, \cdots, \hat{e_{j}}, \cdots, e_{n+1},
e_{n+2}] = \sum\limits_{k=1}^{r}b_{ij}^{k}e_{k};
\end{array}\right.
\end{array}$

\vspace{2mm}\noindent $\begin{array}{ll} (3)~
\left\{\begin{array}{l}
{[}e_{2}, \cdots, e_{n+1}] = e_{1},\\
{[}e_{1}, e_{3}, \cdots, e_{n+1}] = e_{2},\\
{[}e_{1}, \cdots, \hat{e_{i}}, \cdots, \hat{e_{j}}, \cdots, e_{n+1},
e_{n+2}] = \sum\limits_{k=1}^{r}b_{ij}^{k}e_{k};
\end{array}\right.
\end{array}$

\vspace{2mm}\noindent $\begin{array}{ll} (4)~
\left\{\begin{array}{l}
{[}e_{2}, \cdots, e_{n+1}] =\alpha e_{1}+e_{2},\\
{[}e_{1}, e_{3}, \cdots, e_{n+1}] = e_{2},\\
{[}e_{1}, \cdots, \hat{e_{i}}, \cdots, \hat{e_{j}}, \cdots, e_{n+1},
e_{n+2}] = \sum\limits_{k=1}^{r}b_{ij}^{k}e_{k};
\end{array}\right.
\end{array}$

\vspace{2mm}\noindent $\begin{array}{ll} (5)~
\left\{\begin{array}{l}
{[}e_{1}, e_{3}, \cdots, e_{n+1}] =e_{1},\\
{[}e_{2}, e_{3}, \cdots, e_{n+1}] = e_{2},\\
{[}e_{1}, \cdots, \hat{e_{i}}, \cdots, \hat{e_{j}}, \cdots, e_{n+1},
e_{n+2}]=\sum\limits_{k=1}^{r}b_{ij}^{k}e_{k};
\end{array}\right.
\end{array}$

\vspace{2mm}\noindent $\begin{array}{ll} (6)~
\left\{\begin{array}{l}
{[}e_{1}, \cdots, \hat{e_{i}}, \cdots, e_{m}, \cdots, e_{r}, \cdots, e_{n+1}] = e_{i},  ~1\leq i\leq m,3\leq m \leq r-1,\\
{[}e_{1}, \cdots, \hat{e_{i}}, \cdots, \hat{e_{j}}, \cdots, e_{n+1},
e_{n+2}]=\sum\limits_{k=1}^{r}b_{ij}^{k}e_{k};
\end{array}\right.
\end{array}$

\vspace{2mm}\noindent $\begin{array}{ll} (7)~
\left\{\begin{array}{l}
{[}e_{1}, \cdots, \hat{e_{i}}, \cdots, e_{r}, e_{r+1}, \cdots, e_{n+1}] = e_{i},   ~1\leq i\leq r,\\
{[}e_{1}, \cdots, \hat{e_{i}}, \cdots, \hat{e_{j}}, \cdots, e_{n+1},
e_{n+2}]=\sum\limits_{k=1}^{r}b_{ij}^{k}e_{k};
\end{array}\right.
\end{array}$

\vspace{2mm}\noindent where $b_{ij}\in F, ~1\leq i< j\leq n+1$.

Firstly, we study the case (1). For substituting  ${[}e_{2}, \cdots,
e_{n+1}] = e_{1}$ into the other equations  and by the Jacobi
identities on $\{[e_{2}, \cdots, e_{n+1}],$ $e_{2},$ $\cdots, e_{r},
e_{r+1},$ $\cdots, \hat{e_{i}},$ $\cdots,$ $ ~\hat{e_{j}},$
$\cdots,~ e_{n+1},$ $e_{n+2}\}$ for $r+1\leq i < j\leq n+1;$
$\{[e_{2}, \cdots, e_{n+1}], e_{2},$ $\cdots,$ $~\hat{e_{i}},$
$\cdots,$ $e_{r},$ $e_{r+1},$ $~\cdots,$ $~\hat{e_{j}},$ $\cdots,~
e_{n+1},~ e_{n+2}\}$ for $2\leq i\leq r,$ $r+1\leq j\leq n+1;$
$\{[e_{2}, \cdots, e_{n+1}],$ $e_{2},$ $\cdots,$ $~\hat{e_{i}},$
$\cdots,$ $\hat{e_{j}},$ $\cdots,$ $e_{r},$ $e_{r+1},$ $\cdots,$
$e_{n+2}\}$ for $2\leq i < j\leq r$, we get $b_{ij}^k=0$ for $1\leq
k\leq r,$ ~$r+1\leq i < j\leq n+1;$
$b_{ij}^{1}=(-1)^{i-2}b_{1j}^{i},$ $b_{ij}^{2}=\ldots=b_{ij}^{r}=0$
for  $2\leq i\leq r,$ $r+1\leq j\leq n+1;$ and
$b_{ij}^{1}=(-1)^{i-2}b_{1j}^{i}+(-1)^{j-3}b_{1i}^{j},b_{ij}^{2}=\ldots=b_{ij}^{r}=0$
for $2\leq i < j\leq r$ respectively.

Replacing
$e_{n+2}-\sum\limits_{j=2}^{n+1}(-1)^{n+1-j}b_{1j}^{1}e_{j}$ for
$e_{n+2}$, we get the isomorphic form of (1)
$$\begin{array}{ll}
(1)' ~\left\{\begin{array}{l}
{[}e_{2}, \cdots, e_{n+1}] = e_{1},\\
{[}e_{2}, \cdots, \hat{e_{j}}, \cdots, e_{r}, \cdots, e_{n+2}]=
\sum\limits_{k=2}^{r}b_{1j}^{k}e_{k}, 2\leq j \leq r,\\
{[}e_{2}, \cdots, e_{r}, e_{r+1}, \cdots, \hat{e_{j}}, \cdots,
e_{n+1}, e_{n+2}] = \sum\limits_{k=2}^{r}b_{1j}^{k}e_{k},
r+1\leq j \leq n+1,\\
{[}e_{1}, \cdots, \hat{e_{i}}, \cdots, \hat{e_{j}}, \cdots, e_{r},
\cdots, e_{n+2}]=
((-1)^{i-2}b_{1j}^{i}+(-1)^{j-3}b_{1i}^{j})e_{1}, 2\leq i<j \leq r,\\
{[}e_{1}, \cdots, \hat{e_{i}}, \cdots, e_{r}, e_{r+1}, \cdots,
\hat{e_{j}}, \cdots, e_{n+2}] =(-1)^{i}b_{1j}^{i}e_{1}, 2\leq i \leq
r<j\leq n+1.
\end{array}\right.
\end{array}$$

If fixing $e_{n+2}$ in the multiplication of $A$, we get an $(n+2)$-
dimensional $(n-1)$-Lie algebra $A_3=A$ (as vector spaces) with the
product $[x_1, \cdots, x_{n-1}]_3=[x_1, \cdots, x_{n-1}, e_{n+2}]$
for $\forall x_1, \cdots, x_{n-1}\in A_3$, and the multiplication
table in the basis $e_1, \cdots, e_{n+2}$ is as follows

\vspace{2mm}\noindent $\begin{array}{ll} ~\left\{\begin{array}{l}
{[}e_{2}, \cdots, \hat{e_{j}}, \cdots, e_{r}, \cdots, e_{n+1}]_3=
\sum\limits_{k=2}^{r}b_{1j}^{k}e_{k}, 2\leq j \leq r,\\
{[}e_{2}, \cdots, e_{r}, e_{r+1}, \cdots, \hat{e_{j}}, \cdots,
e_{n+1}]_3 = \sum\limits_{k=2}^{r}b_{1j}^{k}e_{k},
r+1\leq j \leq n+1,\\
{[}e_{1}, \cdots, \hat{e_{i}}, \cdots, \hat{e_{j}}, \cdots, e_{r},
\cdots, e_{n+1}]_3=
((-1)^{i}b_{1j}^{i}+(-1)^{j-1}b_{1i}^{j})e_{1}, 2\leq i<j \leq r,\\
{[}e_{1}, \cdots, \hat{e_{i}}, \cdots, e_{r}, e_{r+1}, \cdots,
\hat{e_{j}}, \cdots, e_{n+1}]_3 =(-1)^{i}b_{1j}^{i}e_{1}, 2\leq i
\leq r<j\leq n+1.
\end{array}\right.
\end{array}$

\vspace{2mm} Set $B=Fe_2+\cdots+Fe_{n+1}$. Then $B$ is a subalgebra
of $A_3$ with multiplication table

\vspace{2mm}\noindent $\begin{array}{ll} (1_b)~
\left\{\begin{array}{l} {[}e_{2}, \cdots, \hat{e_{j}}, \cdots,
e_{r}, \cdots, e_{n+1}]_3=
\sum\limits_{k=2}^{r}b_{1j}^{k}e_{k}, ~2\leq j \leq r,\\
{[}e_{2}, \cdots, e_{r}, e_{r+1}, \cdots, \hat{e_{j}}, \cdots,
e_{n+1}]_3=
 \sum\limits_{k=2}^{r}b_{1j}^{k}e_{k}, ~r+1\leq j \leq n+1.\\
\end{array}\right.
\end{array}$

By similar discussions leading to Theorem 3 in [3], we get

\vspace{2mm}\noindent $b_{1j}^{2}=\cdots=b_{1j}^{r}=0, ~r+1\leq j
\leq n+1;(-1)^{i-2}b_{1j}^{i}+(-1)^{j-3}b_{1i}^{j}=0.$

 \vspace{2mm}\noindent Taking a suitable transformation of basis $e_2, \cdots, e_{n+1},$
we get $(1_b)$ isomorphic to

\vspace{2mm}\noindent $\begin{array}{ll}
 \left\{\begin{array}{l}
 {[}e_{3}, \cdots,  e_{n+1}]_3 = e_{2},\\
 \cdots~~~~\cdots~~~~\cdots~~~~\\
 {[}e_{2}, \cdots, \hat{e_i}, \cdots, e_{n+1}]_3 = e_{i}, ~2\leq
i\leq r,\\
  \cdots~~~~\cdots~~~~\cdots~~~~\\
 {[}e_{3}, \cdots,  e_{r-1}, e_{r+1}, \cdots, e_{n+1} ]_3 = e_{r}.
\end{array}\right.
\end{array}$

\vspace{2mm}\noindent Therefore, after taking a suitable $e_1,$ we
get (1) is isomorphic to

\vspace{2mm}\noindent $\begin{array}{ll} (r^{1})~
\left\{\begin{array}{l}
{[}e_{2}, \cdots, e_{n+1}] = e_{1},\\
{[}e_{3}, \cdots, e_{n+2}] = e_{2},\\
 \ldots  \quad  \ldots \quad \ldots \quad \ldots  ,\\
{[}e_{2}, \cdots, \hat{e_i}, \cdots, e_{r}, \cdots, e_{n+2}]=e_{i},\\
 \ldots  \quad  \ldots \quad \ldots \quad \ldots  ,\\
{[}e_{2}, \cdots, e_{r-1}, e_{r+1}, \cdots, e_{n+2}] = e_{r}.
\end{array}\right.
\end{array}$

\vspace{2mm} Secondly, we study the case $(2)$. Replacing $e_{n+2}$
by $e_{n+2}-\sum\limits_{i=1}^{n}(-1)^{n-i}b_{in+1}^{1}e_{i}$
 in (2), we get

\vspace{2mm}\noindent $\begin{array}{ll} (2)'
~\left\{\begin{array}{l}
{[}e_{1}, \cdots, e_{n}] = e_{1},\\
{[}e_{1}, \cdots, \hat{e_{i}}, \cdots, \hat{e_{j}}, \cdots, e_{n},
e_{n+1}, e_{n+2}]=
\sum\limits_{k=1}^{r}b_{ij}^{k}e_{k}, ~1\leq i<j \leq n,\\
{[}e_{1}, \cdots, \hat{e_{i}}, \cdots, e_{n}, e_{n+2}]=
\sum\limits_{k=2}^{r}b_{in+1}^{k}e_{k}, ~1\leq i \leq n.\\
\end{array}\right.
\end{array}$

\vspace{2mm}\noindent Substituting ${[}e_{1},~  \cdots,~ e_{n}] =
e_{1}$ into other equations and by the Jacobi identities on
$\{[e_{1},$ $\cdots,$ $ e_{n}],$ $e_{2}, \cdots, \hat{e_{i}},$
$\cdots,$ $e_{r},$ $e_{r+1},$ $\cdots,$ $e_{n},$ $e_{n+2}\}$ for
$2\leq i \leq r$, $\{[e_{1},\cdots,~ e_{n}],$ $e_{2},$ $\cdots,$
$e_{r},$ $e_{r+1},$ $ \cdots,$ $\hat{e_{i}},$ $\cdots,$ $e_{n},$
$e_{n+2}\}$ for $r+1\leq i\leq n$,
$\{[e_{1},\cdots,e_{r-1},e_{r+1},\cdots, e_{n},e_{n+2}],$ $e_{2},$
$\cdots,$ $~\hat{e_{j}},$ $\cdots,$ $e_{n+1}\}$ for $i=1,2\leq j\leq
n$, $\{[e_{1},\cdots,~ e_{n}],$ $ e_{2},$ $\cdots,$ $\hat{e_{i}},$
$\cdots,$ $~\hat{e_{j}},$ $\cdots,$ $e_{n+1},e_{n+2}\}$ for $2\leq
i<j\leq n,$
 we get $b_{in+1}^{2}=\ldots=b_{in+1}^{r}=0,b_{1n+1}^{i}=0$ for $2\leq i \leq
r$, $b_{in+1}^{2}=\ldots=b_{in+1}^{r}=0$ for $r+1\leq i\leq n,$
$b_{1j}^{r}=0$ for $i=1,2\leq j\leq n,$ $b_{ij}^{r}=0$ for $2\leq
i<j\leq n$ respectively. Then (2) is of the form

\vspace{2mm}\noindent $\begin{array}{ll} \left\{\begin{array}{l}
{[}e_{1}, \cdots, e_{n}] = e_{1},\\
{[}e_{1}, \cdots, \hat{e_{i}}, \cdots, \hat{e_{j}}, \cdots, e_{n+1},
e_{n+2}]=\sum\limits_{k=1}^{r-1}b_{ij}^{k}e_{k}, ~1\leq i<j \leq n.
\end{array}\right.
\end{array}$

\vspace{2mm}\noindent We get $\dim A^{1}=r-1$, this is a
contradiction. Therefore the case (2) is not realized.

By similar arguments to above, we get that cases (3), (4), (5)and
(6) are not realized.

Lastly we study the case $(7)$. For $r+1\leq i<j\leq n+1$, imposing
the Jacobi identities on $\{[e_{2}, \cdots, ~ e_{n+1}],$ $e_{2},$ $
\cdots,$ $e_{r},$ $e_{r+1},$ $\cdots,$ $\hat{e_{i}},$ $\cdots,$ $
~\hat{e_{j}},$ $\cdots,$ $ e_{n+1},$ $ e_{n+2}\}$, we get
$b_{ij}^{k}=0,~ 1\leq k\leq r.$ Then (7) has the form

\vspace{2mm}\noindent $\begin{array}{ll} \left\{\begin{array}{l}
{[}e_{1}, \cdots, \hat{e_{i}}, \cdots, e_{r}, \cdots, e_{n+1}] = e_{i}, ~1\leq i\leq r,\\
{[}e_{1}, \cdots, \hat{e_{i}}, \cdots, \hat{e_{j}}, \cdots, e_{n+1},
e_{n+2}]=\sum\limits_{k=1}^{r}b_{ij}^{k}e_{k}, ~1\leq i \leq r,i<j
\leq n+1.
\end{array}\right.
\end{array}$

For every $i\neq p,$ $ 1\leq i, p\leq r$, substituting ${[}e_{1},$
$\cdots,$ $\hat{e_i},$ $\cdots,$ $e_r, e_{r+1},$ $\cdots,$ $e_{n+1}]
= e_{i}$ into the equation

\vspace{2mm}\noindent
$\sum\limits_{k=1}^{r}b_{ij}^{k}e_{k}={[}e_{1},\cdots,e_{p-1},e_{p+1},\cdots,
 e_{r},\cdots,~\hat{e_{j}},\cdots,e_{n+1},e_{n+2}], ~r+1\leq j\leq n+1,$ we get
$$ b_{pj}^{p}=b_{1j}^{1},~~  b_{pj}^{k}=0 ~~\mbox{for}~ k\neq p,
~1\leq p, ~k \leq r< j\leq n+1.$$

For $2\leq j\leq r$; substituting  ${[}e_{1},$ $\cdots,$
$\hat{e_i},$ $\cdots,$ $e_r, e_{r+1},$ $\cdots,$ $e_{n+1}] = e_{i},
2\leq i \neq j\leq r$ into the equation

\vspace{2mm}\noindent
$\sum\limits_{k=1}^{r}b_{1j}^{k}e_{k}={[}e_{2},\cdots,~\hat{e_{j}},\cdots,e_{r},e_{r+1},\cdots,e_{n+2}],
$

\vspace{2mm}\noindent we get

\vspace{2mm}\noindent $b_{1j}^{k}=0, ~~
b_{1j}^{1}=\ldots=b_{j-1j}^{j-1}
=-b_{jj+1}^{j+1}=\ldots=-b_{jr}^{r}, ~~b_{1j}^{j}=b_{12}^{2} ~~
\mbox{for} ~~2\leq k\neq j\leq r.$

\vspace{2mm} For $2\leq i<j\leq r$, substituting ${[}e_{1},~ \cdots,
\hat{e_p}, \cdots, ~ e_{n+1}]= e_{p}, ~ 1\leq p\leq r $ and $p\neq
i, j$ into the equation

\vspace{2mm}\noindent
$\sum\limits_{k=1}^{r}b_{ij}^{k}e_{k}={[}e_{1},e_{2},\cdots,~\hat{e_{i}},
\cdots,~\hat{e_{j}}, \cdots, e_{r}, \cdots, e_{n+2}],$

\vspace{2mm}\noindent  we get
$$b_{ij}^{k}=0 ~\mbox{for} ~1\leq k\leq r, ~k\neq i, j, ~\mbox{and}~ 2\leq i<j\leq r.$$
Then (7) is isomorphic to

\vspace{2mm}\noindent $\begin{array}{ll}
(7)'~\left\{\begin{array}{l}
{[}e_{1}, \cdots, \hat{e_{i}}, \cdots, e_{r}, e_{r+1}, \cdots, e_{n+1}] = e_{i},~ 1\leq i \leq r,\\
{[}e_{2}, \cdots, \hat{e_{j}}, \cdots, e_{r}, e_{r+1}, \cdots, e_{n+2}]=b_{1j}^{1}e_{1}+b_{12}^{2}e_{j},  ~2\leq j \leq r,\\
{[}e_{1}, e_{2}, \cdots, \hat{e_{i}}, \cdots, \hat{e_{j}}, \cdots,
e_{r}, e_{r+1}, \cdots, e_{n+2}]=
b_{1j}^{1}e_{i}-b_{1i}^{1}e_{j},  ~ 2\leq i<j\leq r,\\
{[}e_{1}, \cdots, \hat{e_{i}}, \cdots, e_{r}, e_{r+1}, \cdots,
\hat{e_{j}}, \cdots, e_{n+1}, e_{n+2}]=b_{1j}^{1}e_{i},~1\leq i \leq
r<j\leq n+1.
\end{array}\right.
\end{array}$

\vspace{2mm}\noindent Replacing  $e_{n+2}$ by
$e_{n+2}-\sum\limits_{j=2}^{n+1}(-1)^{n+1-j}b_{1j}^{1}e_{j}-(-1)^{n-1}b_{12}^{2}e_{1}$
in $(7)'$, we get

\vspace{2mm}\noindent $\begin{array}{ll} (r^{2})~
\left\{\begin{array}{l}
{[}e_{2}, \cdots, e_{n+1}] = e_{1},\\
 \ldots  \quad  \ldots \quad \ldots \quad \ldots  ,\\
{[}e_{1}, \cdots, \hat{e_i}, \cdots, e_{r}, \cdots, e_{n+1}] = e_{i},\\
 \ldots  \quad  \ldots \quad \ldots \quad \ldots  ,\\
 {[}e_{1}, \cdots, e_{r-1}, e_{r+1}, \cdots, e_{n+1}] = e_{r}.
\end{array}\right.
\end{array}$

\vspace{2mm}\noindent It is evident that $(r^1)$ is not isomorphic
to $(r^2)$. \hfill${\Box}$

\vspace{6mm}

 \vspace{2mm}\noindent {\bf
Acknowledgement}: The third author acknowledges the support of the
Australian Research Council.

\vspace{5mm}\noindent{\large References}

\begin{description}

\item{[1]} Y. Nambu, Generalized Hamiltonian Dynamics, {\it Phys. Rev.,} 1973, D7,
                2405-2412
\item{[2]} L. Takhtajan, On foundation of the generalized Nambu mechanics, {\it Commun.
                 Math. Phys.,} 1994, 160, 295-315
\item{[3]}  V. T. Filippov, $n-$Lie algebras,  {\it Sib. Mat.
           Zh.,} 1985, 26(6), 126-140
\item{[4]} J. Bagger and N. Lambert, Gauge symmetry
           and supersymmetry of multiple $M2$-branes, {\it Phys. Rev.,}
           D77, 2008, 065008, arXiv: 0711.0955[hep-th]
\item{[5]} P. Ho, R. Hou and Y. Matsuo, Lie $3$-algebra and multiple
$M_2$-branes, arXiv: 0804. 2110 v2[hep-th]
\item{[6]} P. Ho, M. Chebotar, W. Ke, On skew-symmetric maps on Lie algebras,
{\it Proc. Royal Soc. Edinburgh,} 2003, 133A. 1273-1281
\item{[7]} G, Papadopoulos, $M2$-branes, $3$-Lie algebras and Plucker
          relations, arXiv: 0804. 2662[hep-th]
\item{[8]} D. Alekseevsky and P. Guha, On decomposability
              of Nambu-Poisson Tensor,
              {\it Acta Mathematica Universitatis Comenianae,} 1996, 65, 1-9
\item{[9]} P. Gautheron, Simple facts concerning Nambu
algebras, {\it Commun. Math. Phys.,} 1998, 195 417-34

\item{[10]} P. W. Michor and A. M. Vinogradov n-ary and
associative algebras, {\it Rend. Sem. Mat. Univ. Pol. Torino,} 1996,
53 373-92
\item{[11]} G. Marmo, G. Vilasi and A. M. Vinogradov, The local
structure of n-Poisson and n-Jacobi manifolds, {\it J. Geom. Phys.,}
1998, 25 141-82
\item{[12]}  N. Nakanishi,  On Nambu-Poisson manifolds
             {\it Rev.~Math.~Phys.,} 1998, 10 499-510
\item{[13]}  A. Vinogradov, M. Vinogradov, On
            multiple generalizations of lie algebras and poisson manifolds,
            {\it American Mathematical Society, Contemp. Math.,} 1998, 219, 273-287
\item{[14]}  W. Ling, On the structure of $n-$Lie
                  algebras, Dissertation, {\it University-GHS-Siegen, Siegn,} 1993.
\item{[15]}  A. P. Pozhidaev, Simple quotient algebras and
             subalgebras of Jacobian algebras, {\it Sib. Math. J.,} 1998,
             39(3), 512-517
\item{[16]}  A. P. Pozhidaev,  Two classes of central
             simple n-Lie algebras, {\it Sib. Math. J.,} 1999, 40(6): 1112-1118
\item{[17]}  R. Bai,  X. Wang,  W. Xiao and H. An, The structure of low dimensional
                $n-$Lie  algebras over a field of characteristic $2$,
                         {\it Linear Alg. Appl.,} 2008, 428, 1912-1920
\item {[18]} R. Bai,  D. Meng,  The strong semi-simple
              $n-$Lie algebras, {\it Commun. Alg.,} 2003, 31(11), 5331-5341
\item{[19]}  D.W. Barnes, On $(n+2)$ dimensional $n-$Lie algebras, arXiv: 0704.1892
\item{[20]}   R. Bai,  D. Meng, Representations of strong semisimple $n$-Lie
    algebras, {\it Adv. Math. (China),} 2006, 35(6), 739-746
\item{[21]} Jose Figueroa-Ofarrill, Lorenzian Lie $n$-algebras, arXiv: 0805,
4760v3[math.Rt]

\item{[22]} R. Bai,  G. Song, The classification of six-dimensional 4-Lie algebras,
 {\it J. Phys. A: Math. Theor.} 2009, 42 035207 (17pp)

\item{[23]} W. A. de Graaf,  Classification of solvable Lie algebras,
               Experimental Mathematics, 2005, 14: 15-25

\end{description}
\end{document}